\documentclass{jpsj3}
\usepackage{color}

\def\runtitle{
Intrinsic Angular Momentum and Intrinsic Magnetic Moment of Chiral Superconductor
}
\def\runauthor{
Atsushi {\sc Tsuruta}, Susumu {\sc Yukawa} and Kazumasa {\sc Miyake}
}

\title{
Intrinsic Angular Momentum and Intrinsic Magnetic Moment 
of Chiral Superconductor on Two-Dimensional Square Lattice
}
\author{
Atsushi {\sc Tsuruta}$^{1}$, Susumu {\sc Yukawa}$^{1}$, 
and Kazumasa {\sc Miyake}$^{2}$
}
\inst
{
$^{1}$Division of Materials Physics, Department of Materials Engineering Science, \\
Graduate School of Engineering Science, Osaka University, Toyonaka 560-8531, Japan\\
$^{2}$Toyota Physical and Chemical Research Institute, Nagakute, Aichi 480-1192, Japan
}

\recdate
{
June 23, 2015
}

\abst
{
The intrinsic magnetic moment (IMM) and intrinsic angular momentum (IAM) of a chiral superconductor 
with $p$-wave symmetry on a two-dimensional square lattice are discussed on the basis of the 
Bogoliubov-de Gennes equation.  The the IMM and IAM are shown to be on the order of 
$\mu_{\rm B}N$ and $\hbar N$, respectively, $N$ being the total number of particles, 
without an extra factor 
$(T_{\rm c}/T_{\rm F})^{\gamma}$ ($\gamma=1,2$), and parallel to the pair angular momentum.  
They arise from the current in the surface layer with a width on the order of the coherence length 
$\xi_{0}$, the size of Cooper pairs.  However, in a single-band model, they are 
considerably canceled by the contribution from the Meissner surface current in a layer with 
the width of the penetration depth $\lambda$, making it difficult to observe them experimentally. 
In the case of multi-band metals with both electron-like and hole-like bands, however, 
considerable cancellation still occurs for the IMM but not for the IAM, 
making it possible to observe the IAM selectively because the effect of the Meissner current 
becomes less important.  As an example of a multi-band metal, the case of 
the spin-triplet chiral superconductor Sr$_2$RuO$_4$ is discussed and 
experiments for observing the IAM are proposed.   
}

\begin{document}
\sloppy
\maketitle
\section{Introduction}
\subsection{Issues concerning intrinsic angular momentum in $^3${\rm He A}-phase}
The size of the intrinsic angular momentum (IAM) $L_{\rm in}$ has been a major issue 
since the middle of the '70s, when the size of the IAM in the A-phase of superfluid $^{3}$He 
was keenly discussed. 
Namely, the issue is the value of the exponent $\gamma$ when the IAM in the ground state is 
expressed as $L_{\rm in}=(N\hbar/2)(T_{\rm c}/T_{\rm F})^{\gamma}$ in the case 
where the ${\bf {\hat \ell}}$-vector is uniform in space.  
This problem was first addressed by Anderson and Morel in their seminal paper~\cite{AndersonMorel} 
discussing a possible state of superfluidity of liquid $^3$He, who stated that $\gamma=1$.  
This estimation is based on the idea that the IAM is sustained by a pair condensate 
with angular momentum $\hbar$ so that the IAM is proportional to the pair amplitude 
($\propto T_{\rm c}/T_{\rm F}$), as is clearly elucidated by Leggett in a review article 
on superfluid $^3$He~\cite{Leggett}.  After the discovery of the superfluidity of $^3$He, 
a bunch of theoretical works were performed on the size of the IAM by microscopic calculations 
based on the {\bf k}-space approach, which predicted 
$\gamma=2$~\cite{Wolfle, Cross1,Volovik,Combescot,Balatsky}. 
This result seems to have been interpreted such that the IAM is formed by two particles inside the 
thin surface layer in a $k$-space near the Fermi surface whose width is on the order of 
the superfluid gap $\Delta$, implying that the number of particles forming the IAM is 
of the order of $N\times(\Delta/E_{\rm F})^{2}$ or $N\times(T_{\rm c}/T_{\rm F})^{2}$.  

On the other hand, in 1976, Ishikawa proposed an alternative idea~\cite{Ishikawa} 
that the IAM in the A-phase is $L_{\rm in}=(N\hbar/2)$ in the ground state if we adopt the wave 
function proposed by BCS, whose property was discussed by Ambegaokar in the review article 
appearing in the textbook on superconductivity edited by Parks~\cite{Ambegaokar}.  
The point is that the orbital-part of the many particle 
ground state $\Psi({\bf r}_1,{\bf r}_2,{\bf r}_3,{\bf r}_4,\cdots,{\bf r}_{N-1},{\bf r}_N)$ 
($N$ being an even number) 
is given by the product of the wave function of the Cooper pair $\varphi({\bf r},{\bf r}^{\prime})$ 
which is the eigenstate of the relative angular momentum $\ell_{z}=\hbar$ concerning 
$({\bf r}-{\bf r}^{\prime})$.  Namely, 
\begin{equation}
\Psi({\bf r}_1,{\bf r}_2,{\bf r}_3,{\bf r}_4,\cdots,{\bf r}_{N-1},{\bf r}_N)=
{\cal A}\left[\varphi({\bf r}_1,{\bf r}_2)
\varphi({\bf r}_3,{\bf r}_4)\cdots\varphi({\bf r}_{N-1},{\bf r}_N)\right],
\label{eq:1}
\end{equation}
where ${\cal A}$ indicates the anti-symmetrization with respect to all the spatial coordinates, 
and the spin coordinates are discarded as irrelevant.  
Indeed, Ambegaokar showed explicitly that this form of the wave function can be transformed into 
the wave function proposed by BCS if the Fourier component of the wave function of the Cooper pair 
$\varphi({\bf r},{\bf r}^{\prime})$ is well defined with respect to the relative coordinate 
$({\bf r}-{\bf r}^{\prime})$.  The issue at that stage was whether the last condition is 
satisfied or not.  Indeed, the Fourier component of 
$\varphi({\bf r}-{\bf r}^{\prime})$, $\varphi_{\bf k}$, 
is given by 
\begin{equation}
\varphi_{\bf k}={v_{\bf k}\over u_{\bf k}},
\label{eq:2}
\end{equation}
where the coefficient $u_{\bf k}$ expressing the coherence of the BCS state vanishes outside 
the thin region around the Fermi surface of the width of cut-off $\varepsilon_{\rm c}$ so long as the 
weak-coupling theory is applied. However, the weak-coupling theory might discard a crucial 
effect on the coherence between the region near the Fermi surface and the core region of the 
Fermi sphere.  Namely, the issue is to what extent the coherence of the Cooper pair condensate 
is maintained into the core of the Fermi sphere in a real situation.  This seems to be a fundamental 
question regarding how to understand the Fermi superfluid state, including superconductivity.  

It is instructive to remember the discussions of Bogoliubov {\it et al}\/.~\cite{Bogoliubov} 
and Anderson and Morel,~\cite{AndersonMorel} who estimated the effect of the Coulomb repulsion 
on the Cooper pair formation.  They solved the gap equation with the phonon-mediated attractive 
interaction near the Fermi surface with a width on the order of the Debye energy $\omega_{\rm D}$ 
and the Coulomb repulsive interaction acting in the whole $k$-space, and they obtained 
a superconducting 
gap that is finite not only in the thin surface around the Fermi surface but also in the whole 
$k$-space including the core of the Fermi sphere.  This was crucial to understanding the systematic 
deviation of the index $\alpha$ for the isotope effect from the canonical value 
$\alpha=1/2$~\cite{deGennes}.  
Therefore, it is a real effect that the coherence of the Cooper pair extends to the 
core of the Fermi sphere.  In this sense, a difficulty posed for the idea of Ishikawa was safely 
avoided~\cite{IMU,Kita}.  
Adopting the wave function in Eq.\ (\ref{eq:1}), McClure and Takagi showed~\cite{McClure} 
that the result 
$L_{\rm in}=N\hbar/2$ holds more generally within the manifold of the ground state with an 
axially symmetric configuration of the ${\bf {\hat \ell}}$-vector, such as the  
Mermin-Ho~\cite{MerminHo} and Anderson-Toulouse~\cite{AndersonToulouse} textures.  
Namely, the state with a uniform configuration of the ${\bf {\hat \ell}}$-vector is 
adiabatically continued from such axially symmetric states.  

A related issue concerned the structure of the supercurrent ${\bf j}_{\rm s}$ in the A-phase of 
superfluid $^{3}$He.  By the symmetry argument, ${\bf j}_{\rm s}$ is expressed as 
\begin{equation}
{\bf j}_{\rm s}={\tilde \rho}_{\rm s}{\bf v}_{\rm s}+{\tilde C}(\nabla\times{\bf {\hat \ell}})
+{1\over 2}\nabla\times\left(L_{\rm in}{\bf {\hat \ell}}\right),
\label{eq:3}
\end{equation}
where ${\tilde \rho}_{\rm s}$ is the superfluid density tensor and the tensor ${\tilde C}$ has 
the following form in general: 
\begin{equation}
C_{ij}=C\delta_{ij}-C_{0}{\hat \ell}_{i}{\hat \ell}_{j}.
\label{eq:4}
\end{equation}
The issue concerned the form of ${\tilde C}$ together with the size of $L_{\rm in}$.  
According to microscopic calculations based on the ${\bf k}$-space representation, 
$C_{ij}$ is given as $C_{ij}=C(T)(\delta_{ij}-2{\hat \ell}_{i}{\hat \ell}_{j})$ and 
$C(T=0)=\hbar n/4$, where $n$ is the number density, and 
$L_{\rm in}\sim \hbar n_{\rm s}(T_{\rm c}/T_{\rm F})^{2}$~\cite{Cross1,Cross2} or 
$L_{\rm in}=0$~\cite{Wolfle}, where $n_{\rm s}$ is the superfluid number density.  
On the other hand, the supercurrent density at $T=0$ is given by 
\begin{equation}
{\bf j}_{\rm s}=\rho_{\rm s}{\bf v}_{\rm s}+
{1\over 2}\nabla\times\left({1\over 2}\hbar n_{\rm s}{\bf {\hat \ell}}\right),
\label{eq:5}
\end{equation}
according to a modified wave function based on Eq.\ (\ref{eq:1}), which is generalized 
so as to take into 
account a gradual variation of the center of mass coordinate in 
$\varphi({\bf r},{\bf r}^{\prime})$~\cite{IMU}. 
This supercurrent is in a form parallel to that appearing in the electromagnetism in 
materials~\cite{Purcell}, in which the electric current ${\bf j}_{\rm M}$ induced by a variation of 
the magnetization density ${\bf M}$ is given by 
\begin{equation}
{\bf j}_{\rm M}={1\over \mu_{0}}\nabla\times{\bf M}. 
\label{eq:6}
\end{equation}
Indeed, if one remembers the relation ${\bf M}=(e\hbar/2m){\bf L}$, ${\bf L}$ being the 
orbital angular momentum, and that ${\bf j}_{\rm s}$ corresponds to $(m/e){\bf j}_{\rm M}$, 
the second term in Eq.\ (\ref{eq:5}) is precisely 
Eq.\ (\ref{eq:6}).  On the other hand, Mermin and Muzikar~\cite{MerminMuzikar} claimed that 
the calculation in Ref.\ \citen{IMU} misses a subtle singularity of the wave function 
$\varphi({\bf r},{\bf r}^{\prime})$ when it is calculated in the ${\bf k}$-space representation, 
leading to the expression for the supercurrent 
\begin{equation}
{\bf j}_{\rm s}=\rho{\bf v}_{\rm s}+
{1\over 4}\hbar (\nabla n_{\rm s}) \times {\bf {\hat \ell}}.  
\label{eq:7}
\end{equation}
This form was also derived by Nagai on the basis of kinetic theory~\cite{Nagai}. 
The physical meaning of the difference between Eqs.\ (\ref{eq:5}) and (\ref{eq:7}) 
has been discussed by Volovik from a general viewpoint.~\cite{Volovik2}
In any case, it is crucial that both expressions, Eqs.\ (\ref{eq:5}) and (\ref{eq:7}), include 
the term proportional to $(\nabla n_{\rm s})\times{\bf {\hat \ell}}$ that makes an essential 
contribution to 
the IAM through the surface current when ${\bf {\hat \ell}}$ is parallel to the surface where 
the number density $n$ vanishes abruptly.   

This fact leads to the following physical picture of the IAM in a uniform configuration of the 
${\hat \ell}$-vector.  In the region at a distance from the system boundary of more than the 
coherence length $\xi_{0}$ of the Cooper pairs, the relative motions of the Cooper pairs around 
the ${\hat \ell}$-vector cancel with each other, resulting in no contribution to the 
IAM.  On the other hand, in a thin region with a width on the order of $\xi_{0}$ near the system 
boundary, its cancellation is incomplete, giving rise to the surface current that is the 
main origin of the IAM.  This physical picture is exactly the same as the picture for an 
electric current induced by the (classical) orbital magnetism in materials.  Precisely speaking, 
it is not self-evident that Eq.\ (\ref{eq:5}) or (\ref{eq:7}) cannot be applied 
near a system boundary where the number density changes abruptly, because they were 
derived on the assumption that the spatial variations of physical quantities 
are gradual compared with the length scale characterizing the physics, i.e., the coherence length 
$\xi_{0}$.  

Recently, this problem has been revived in the context of the topological effect 
associated with the surface state of a chiral superfluid or superconductivity.  According to 
detailed calculations that take into account the microscopic structure of the surface state with 
spatial variation on the order of $\xi_{0}$, the result derived from the supercurrent, given 
by Eq.\ (\ref{eq:5}) or (\ref{eq:7}), is essentially correct~\cite{Tsutsumi,Nagato}.  
On the other hand, these calculations show that the IAM at finite temperatures is given by 
\begin{equation}
L_{\rm in}(T)={N_{\rm s}^{\parallel}(T)\hbar\over 2},
\label{eq:8}
\end{equation} 
where $N_{\rm s}^{\parallel}(T)$ is the superfluid number parallel to the 
${\hat \ell}$-vector~\cite{Tsutsumi,Nagato,Sauls}.  This result cannot be easily understood 
intuitively, but seems to be much more involved than a naive physical picture.  
In particular, it is difficult to find the physical reason why $N^{\parallel}$ appears in a 
two-dimensional system when the ${\hat \ell}$-vector is perpendicular to the two-dimensional plane.  
Indeed, it should be compared with the result obtained by a calculation using a 
cylindrical representation for one-particle states, in which $L_{\rm in}(T)$ is 
given by
\begin{equation}
L_{\rm in}(T)={N_{\rm s}^{\perp}(T)\hbar\over 2},
\label{eq:9}
\end{equation} 
where $N_{\rm s}^{\perp}(T)$ is the superfluid number perpendicular to the 
${\hat \ell}$-vector.~\cite{Miyake,Tada}   

Regarding an experiment for detecting the IAM in the $^3$He-A phase, NMR measurement 
of the structural change in the Mermin-Ho texture in a rotating cryostat was proposed by 
Takagi~\cite{Takagi}.  
This experiment has been performed at the Institute for Solid State Physics  
(ISSP) of the University of Tokyo and suggested that the size of the IAM is on the order of 
$N_{\rm s}\hbar/2$~\cite{IshikawaOsamu}.     

\subsection{Issues concerning intrinsic magnetic moment in {\rm Sr}$_2${\rm RuO}$_4$}
In the past decade, the problem concerning the IAM has been revived as that of the 
intrinsic magnetic moment (IMM) in the spin-triplet chiral superconductor Sr$_2$RuO$_4$~\cite{Maeno}, 
in which the orbital part of the superconducting gap has been identified as 
$\Delta_{\bf k}=\Delta(\sin, k_{x}a+{\rm i}\sin\, k_{y}a)$, 
$a$ being the lattice constant of the two-dimensional lattice.  This is also supported by a 
muon-spin-resonance ($\mu$SR) measurement showing the breaking of time-reversal symmetry\cite{muSR}.  
Information on the $k$-representation of $\Delta_{k}$ is obtained from the 
temperature dependence of the specific heat (under a magnetic field) and theoretical investigations 
that suggest the importance of short-range ferromagnetic correlations among 
quasiparticles~\cite{MiyakeNarikiyo,Hoshihara,Yoshioka}.  
If the IAM $L_{\rm in}$ is on the order of $N_{\rm s}\hbar/2$ and the gyromagnetic ratio is given by 
$(-e/2m)$, with $e\,$($>0$) being the elementary charge, as in the classical case, 
the IMM density $M_{\rm in}$ is estimated as 
\begin{equation}
M_{\rm in}\simeq -{n_{\rm s}\over 2}\mu_{0}{m\over m_{\rm band}^{\rm occ}}\mu_{\rm B},
\label{eq:10}
\end{equation} 
where $n_{\rm s}\equiv N_{\rm s}/V$, $\mu_{0}=4\pi\times 10^{-7}\,$H$\cdot$m$^{-1}$ is the magnetic 
permeability of vacuum, $\mu_{\rm B}=e\hbar/2m$ is the Bohr magneton, and 
$m_{\rm band}^{\rm occ}$ is the harmonic average of the band mass of electrons over the 
occupied state in the Brillouin zone {discussed in Appendix\ A and 
should be distinguished from the effective mass averaged over the Fermi surface 
discussed in Appendix\ B. 
}.  
Then, the magnetic flux density $B_{\rm in}$ 
without the external magnetic field $H$ is given by $M_{\rm in}$ because the relation 
$B=M+\mu_{0}H$ holds by definition~\cite{Purcell}.  The electron number density $n$ of the 
$\gamma$-band, which is the electron-like band, in Sr$_2$RuO$_4$, is estimated as 
\begin{equation}
n={1\over abc},
\label{eq:11}
\end{equation}
where $a=b=3.9\times 10^{-10}$ m and $c=(12.7/2)\times 10^{-10}$ m are the lengths of an edge of 
the primitive cell of Sr$_2$RuO$_4$ along the $a$, $b$ and $c$ directions, 
respectively~\cite{Mackenzie}.  The magnetization density $M_{\rm in}$ is given by the relation
\begin{equation}
M_{\rm in}=-\mu_{0}{e\over 2m_{\rm band}^{\rm occ}}L_{\rm in}=
-{\hbar \over 2}n\mu_{0}{e\over 2m_{\rm band}^{\rm occ}},
\label{eq:12}
\end{equation} 
where $m_{\rm band}^{\rm occ}\simeq 2.9\,m$ is the effective mass of the $\gamma$-band of 
Sr$_2$RuO$_4$~\cite{Mackenzie}.  
Therefore, the intrinsic magnetic flux density $B_{\rm in}$ is estimated 
as
\begin{eqnarray}
& &B_{\rm in}=-{10^{-30}\over abc}\,{m\over m_{\rm band}^{\rm occ}}\times 5.8\, {\rm T}
\nonumber
\\
& &\qquad
\simeq -2.1\times 10^{-2}\, {\rm T}=-2.1\times 10^{2}\, {\rm G}.
\label{eq:13}
\end{eqnarray}
This value is larger than the ``observed" lower critical field 
$B^{\rm obs}_{{\rm c}1}=5.0\times10^{-3}\,{\rm T}$ for 
Sr$_2$RuO$_4$~\cite{Akima}.  Therefore, at first sight, this intrinsic magnetic flux density 
$B_{\rm in}$ will not be completely screened by the Meissner current.  
However, since Sr$_2$RuO$_4$ has two other bands, a hole-like $\alpha$-band and an 
electron-like $\beta$-band, considerable cancellation in the IMM is expected among 
the electron-like $\beta$- and the $\gamma$-bands and the hole-like $\alpha$-band, 
as discussed in Sect.\ 5. 

Another issue is whether the IAM is also cancelled by the orbital angular 
momentum arising from the Meissner current moving through a thin surface with the width of the 
penetration depth $\lambda$ 
near the boundary of the system.  If this cancellation is incomplete, the remnant IAM may be 
detected by the Richardson-Einstein-de Haas effect~\cite{Richardson,Einstein}. 
    
\subsection{Purpose and organization of the present paper}
\begin{figure}[h]
\begin{center}
\rotatebox{0}{\includegraphics[width=0.7\linewidth]{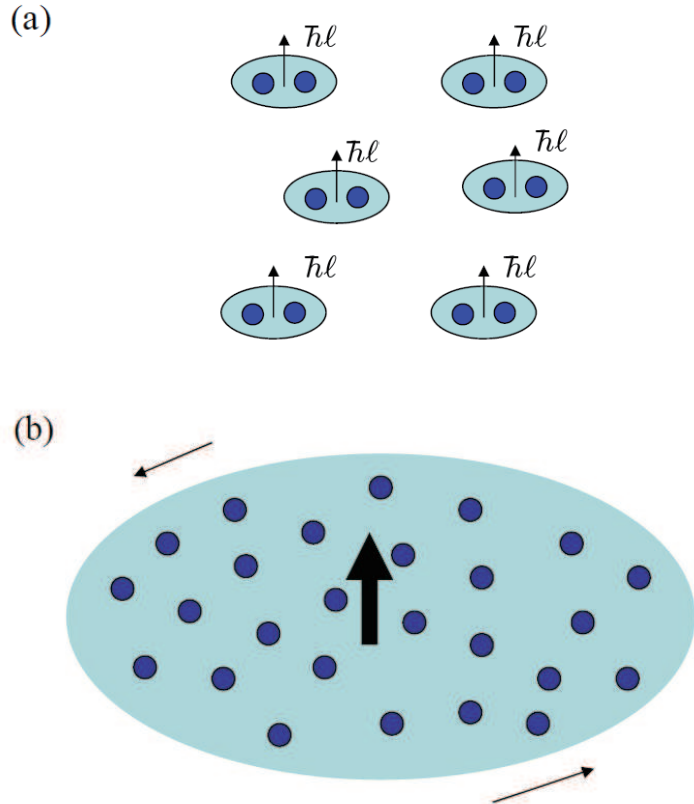}}
\caption{(a) Picture of strong-coupling (BEC) limit, where diatomic molecules with angular momentum 
$\hbar {\hat \ell}$ are in the state of Bose-Einstein condensation. (b) Picture of weak coupling 
(BCS) limit, where the angular momentum is distributed on the Cooper pair condensation.    
}
\label{Fig:1}
\end{center}
\end{figure}

In the ``rotationally symmetric" system, in which the angular momentum is a conserved quantity, 
the result for the IAM, ${\bf L}_{\rm in}=(N\hbar/2){\hat \ell}$, is almost self-evident from  
the viewpoint of the BEC-BCS crossover or of the adiabatic continuation.  
In the BEC limit, each diatomic 
molecule has angular momentum $\hbar {\hat \ell}$ as shown in Fig.\ \ref{Fig:1}(a). 
Therefore, the IAM is given by ${\bf L}_{\rm in}=(N\hbar/2){\hat \ell}$, where 
the number of diatomic molecules is $N/2$.  
So long as the pairing interaction has rotationally symmetry, the value of 
$L_{\rm in}$, which is a conserved quantity, should not change even if the pairing interaction 
is weakened to approach the BCS limit in which ``molecules" overlap each other as shown 
in Fig.\ \ref{Fig:1}(b).  
A subtlety is that the gap amplitude $\Delta$ of a Cooper pair has a very weak singularity. 
Namely, $d^{2}\Delta/d\mu^{2}$ has a discontinuity as the chemical potential $\mu$ passes 
through the bottom of the quasiparticle band~\cite{Randeria}.  

On the other hand, in a lattice system in which the rotation symmetry is broken, 
the problem is not so trivial.  Therefore, it is necessary to explicitly investigate the problem  
for a specific lattice model.  One of the purposes of this paper is to investigate how the 
results for the IAM of the $^3$He-A phase in three-dimensional free space are modified in the case of 
a chiral superconductor on a two-dimensional square lattice, which simulates Sr$_2$RuO$_4$. 
Another purpose is to investigate, using this two-dimensional model, to what extent the IMM is 
screened by the Meissner effect and the IAM is lessened by the Meissner current.  On the basis 
of our results, 
it is discussed how the IMM and IAM are observed in Sr$_2$RuO$_4$.  

The organization of the present paper is as follows.  In Sect.\ 2, we introduce the model 
on a square lattice 
with an attractive interaction between nearest-neighbor sites and discuss a formulation for 
explicit calculations.  In Sect.\ 3, the results for the IAM and IMM are shown.   
In Sect.\ 4, the effect of the Meissner current on the size of the IMM and IAM is discussed.  
Finally, in Sect.\ 5, we propose how to observe the IMM and IAM of the two dimensional chiral 
superconductor Sr$_2$RuO$_4$.   

\section{Chiral Superconductor on Square Lattice}
\subsection{Model Hamiltonian}
In order to study the problem of the IAM and IMM of a chiral superconductor on a 
two-dimensional lattice, a model of Sr$_2$RuO$_4$, we start with the following Hamiltonian: 
\begin{equation}
{\cal H}=-\mu\sum_{i\sigma}c^{\dagger}_{i\sigma}c_{i\sigma}
-t\sum_{\langle i,j\rangle\sigma}c^{\dagger}_{i\sigma}c_{j\sigma}
-{V\over 2}\sum_{\langle i,j\rangle\sigma}
c^{\dagger}_{j\sigma}c^{\dagger}_{i{\bar \sigma}}c_{i{\bar \sigma}}c_{j\sigma},
\label{eq:CS1}
\end{equation}
where $\mu$, $t$, and $V$ are the chemical potential, the transfer integral between nearest-neighbor 
(n.n.) sites of the square lattice, and the attractive interaction between electrons at n.n. sites, 
respectively, and $c^{\dagger}_{i\sigma}$ ($c_{i\sigma}$) is the creation 
(annihilation) operator of an electron at the $i$-th site with spin component $\sigma$ ($=\uparrow$ 
or $\downarrow$).  The symbol $\langle i,j\rangle$ indicates that the summation is taken over 
the n.n. sites.  
Here, we consider the spin-triplet pairing with $S_{z}=0$ and introduce a superconducting gap 
$\Delta_{ij}$ in the spin-triplet manifold as 
\begin{equation}
\Delta_{ij}={V\over 2}\langle c_{i\uparrow}c_{j\downarrow}+
c_{i\downarrow}c_{j\uparrow}\rangle, 
\label{eq:CS2}
\end{equation} 
where $\langle\cdots\rangle$ means the average by the mean-field Hamiltonian 
$H_{\rm mf}$ given as 
\begin{equation}
{\cal H}_{\rm mf}=-\mu\sum_{i\sigma}c^{\dagger}_{i\sigma}c_{i\sigma}
-t\sum_{\langle i,j\rangle\sigma}c^{\dagger}_{i\sigma}c_{j\sigma}
+\sum_{\langle i,j\rangle}\left\{\left[\Delta_{ij}
(c^{\dagger}_{j\uparrow}c^{\dagger}_{i\downarrow}+c^{\dagger}_{j\downarrow}c^{\dagger}_{i\uparrow})
+{\rm h.c.}\right]-\frac{|\Delta_{ij}|^{2}}{V}\right\}.
\label{eq:CS3}
\end{equation}
Here the gap $\Delta_{ij}$ depends on lattice sites $i$ and $j$ in general, and its dependence 
is determined self-consistently by solving the (lattice version of the) 
Bogoliubov-de Gennes equation  
together with Eq.\ (\ref{eq:CS2})~\cite{Onishi}.  
The gap $\Delta_{ij}$ is odd with respect to the interchange of $i\rightleftharpoons j$: 
\begin{equation}
\Delta_{ij}=-\Delta_{ji},
\label{eq:CS2a}
\end{equation}
which results in the odd-parity pairing.  
Note that, in the case of a uniform system without a boundary, the most stable gap among those given 
by Eq.\ (\ref{eq:CS2}) is expressed in a wave-vector 
representation as~\cite{MiyakeNarikiyo}
\begin{equation}
\Delta_{\bf k}=\Delta(\sin\,k_{x}a+{\rm i}\sin\,k_{y}a), 
\label{eq:CS4}
\end{equation}
where $a$ is the lattice constant.  

\subsection{Orbital magnetization and angular momentum in band picture}
In order to take into account the effect of the magnetic field ${\bf B}$, we adopt the 
following way of 
giving the Peierls phase the transfer integral $t_{ij}$ between electrons at the $i$-th and 
$j$-th sites 
\begin{equation}
{\tilde t}_{ij}(B)=t_{ij}\exp\left[{\rm i}\frac{e}{\hbar}
\int_{{\bf r}_i}^{{\bf r}_j}{\bf A}({\bf r})\cdot d{\bf r}\right], 
\label{eq:CS5}
\end{equation}
where ${\bf A}$ is the vector potential giving the magnetic field as 
${\bf B}=\nabla\times {\bf A}$ and 
the contour integral is performed along the line connecting the two sites.   
Then, the band-energy part ${\cal H}_{\rm band}$ of the Hamiltonian is expressed as 
\begin{equation}
{\cal H}_{\rm band}=
-\sum_{\langle i,j\rangle\sigma}{\tilde t}_{ij}(B)\,c^{\dagger}_{i\sigma}c_{j\sigma}.
\label{eq:CS6}
\end{equation}
Adopting the gauge of the vector potential such that 
${\bf A}=({\bf B}\times{\bf r})/2$, ${\cal H}_{\rm band}$ is reduced to the following form: 
\begin{eqnarray}
 &{\cal H}_{\rm band}=-t\sum_{i}
\left[\exp\left({\rm i}{eB\over 2\hbar}y_{i}a\right) c^{\dagger}_{(i_{x}+1,i_{y})}c_{(i_{x},i_{y})}
+\exp\left(-{\rm i}{eB\over 2\hbar}y_{i}a\right) c^{\dagger}_{(i_{x},i_{y})}c_{(i_{x}+1,i_{y})}\right]
\nonumber
\\
 &\qquad\qquad
-t\sum_{i
}\left[\exp\left({\rm i}{eB\over 2\hbar}y_{i}a\right) c^{\dagger}_{(i_{x},i_{y})}c_{(i_{x}-1,i_{y})}
+\exp\left(-{\rm i}{eB\over 2\hbar}y_{i}a\right) c^{\dagger}_{(i_{x}-1,i_{y})}c_{(i_{x},i_{y})}\right]
\nonumber
\\
 &\qquad\qquad
+t\sum_{i}
\left[\exp\left({\rm i}{eB\over 2\hbar}x_{i}a\right) c^{\dagger}_{(i_{x},i_{y}+1)}c_{(i_{x},i_{y})}
+\exp\left(-{\rm i}{eB\over 2\hbar}x_{i}a\right) c^{\dagger}_{(i_{x},i_{y})}c_{(i_{x},i_{y}+1)}\right]
\nonumber
\\
 &\qquad\qquad
+t\sum_{i}
\left[\exp\left({\rm i}{eB\over 2\hbar}x_{i}a\right) c^{\dagger}_{(i_{x},i_{y})}c_{(i_{x},i_{y}-1)}
+\exp\left(-{\rm i}{eB\over 2\hbar}x_{i}a\right) c^{\dagger}_{(i_{x},i_{y}-1)}c_{(i_{x},i_{y})}\right],
\label{eq:CS7}
\end{eqnarray}
where the spin coordinate is abbreviated for concise presentation and 
the $i$-th position of the lattice 
is designated by the two-dimensional representation $(i_{x},i_{y})$.   
In deriving Eq.\ (\ref{eq:CS7}), 
the contour integral Eq.\ (\ref{eq:CS5}) along the $x$-direction has been approximated by the 
trapezoidal rule as 
\begin{equation}
\int_{x_i-a}^{x_i}{A}_x(x,y_{i}) dx\simeq A_{x}\left(x_{i}-{a\over 2},y_{i}\right)a 
\label{eq:CS5b}
\end{equation}
and 
\begin{equation}
\int_{x_i}^{x_i+a}{A}_x(x,y_{i}) dx\simeq A_{x}\left(x_{i}+{a\over 2},y_{i}\right)a. 
\label{eq:CS5c}
\end{equation}
A similar approximation is adopted for the integral along the $y$-direction.  

Then, the operator of the magnetization $M$ of the system (parallel to ${\bf B}$, $z$-component) is 
given, in the limit $B\to 0$, as 
\begin{equation}
\frac{M_{z}}{\mu_{0}}=-\left({\partial {\cal H}\over \partial B}\right)_{B=0}
=-\left({\partial {\cal H}_{\rm band}\over \partial B}\right)_{B=0}
={ta^{2}\over \hbar^{2}}\,(-e)\sum_{i}({\bf r}_{i}\times{\bf p}_{i})_{z},
\label{eq:CS8}
\end{equation}
where the ``momentum" operator ${\bf p}_{i}$ at the $i$-th site is defined by 
\begin{eqnarray}
&p_{xi}\equiv {-{\rm i}\over 2}{\hbar\over a}\sum_{\sigma}
\left[ (c^{\dagger}_{(i_{x}+1,i_{y}){\sigma}}-c^{\dagger}_{(i_{x}-1,i_{y}){\sigma}})c_{(i_{x},i_{y}){\sigma}}
-c^{\dagger}_{(i_{x},i_{y}){\sigma}}(c_{(i_{x}+1,i_{y}){\sigma}}-c_{(i_{x}-1,i_{y}){\sigma}})\right]
\nonumber
\\
&p_{yi}\equiv {-{\rm i}\over 2}{\hbar\over a}\sum_{\sigma}
\left[ (c^{\dagger}_{(i_{x},i_{y}+1){\sigma}}-c^{\dagger}_{(i_{x},i_{y}-1){\sigma}})c_{(i_{x},i_{y}){\sigma}}
-c^{\dagger}_{(i_{x},i_{y}){\sigma}}(c_{(i_{x},i_{y}+1){\sigma}}-c_{(i_{x},i_{y}-1){\sigma}})\right].
\label{eq:CS9}
\end{eqnarray}
If we introduce the band mass $m_{\rm b}$ at the zone boundary, say at the $\Gamma$-point, as 
\begin{equation}
{ta^{2}\over  \hbar^{2}}\equiv{1\over 2m_{\rm b}},
\label{eq:CS10}
\end{equation}
the magnetization operator $M_{z}$ [Eq.\ (\ref{eq:CS8})] is reduced to a band version of the 
conventional form with the gyromagnetic ratio $(-e/2m_{\rm b})$: 
\begin{equation}
M_{z}=\mu_{0}{(-e)\over 2m_{\rm b}}\sum_{i}({\bf r}_{i}\times{\bf p}_{i})_{z}.
\label{eq:CS11}
\end{equation}
This implies that the definition of the orbital angular momentum ${\bf L}$,  
\begin{equation}
{\bf L}\equiv \sum_{i}({\bf r}_{i}\times{\bf p}_{i}),
\label{eq:CS12}
\end{equation} 
is a valid and natural one.  The definition of $m_{\rm b}$ [Eq.\ (\ref{eq:CS10})] corresponds to 
the free-electron-like dispersion of tight-binding dispersion around the 
$\Gamma$-point, $(k_{x},k_{y})=(0,0)$.  
Namely, 
\begin{equation}
\epsilon_{k}=-2t(\cos\,k_{x}a+\cos\,k_{y}a)\simeq -4t+ta^{2}(k_{x}^{2}+k_{y}^{2})+\cdots\,.
\label{eq:CS13}
\end{equation} 
Using this dispersion, $m_{\rm band}^{\rm occ}$ in Eqs.\ (\ref{eq:10}), (\ref{eq:12}), and 
(\ref{eq:13}) is estimated as $m_{\rm band}^{\rm occ}=(\pi^{2}/4)m_{\rm b}$ in the half-filled case, 
as shown in Appendix A.  

\section{Results for IAM and IMM}
An explicit form of the Bogoliubov-de Gennes equation for the mean-field Hamiltonian 
[Eq.\ (15)]  
with the superconducting gap of $S_{z}=0$ [Eq.\ (\ref{eq:CS4})] is given by~\cite{deGennes2}
\begin{eqnarray}
& &\varepsilon\, u_{i}=-\mu \,u_{i}-t\,u_{j}+\sum_{\langle j,i\rangle}\Delta_{ij}v_{j},
\label{eq:R1}
\\
& &\varepsilon\, v_{i}=\mu \,v_{i}+t\,v_{j}+\sum_{\langle j,i\rangle}\Delta_{ij}^{*}u_{j},
\label{eq:R2}
\end{eqnarray}
where $\langle j,i\rangle$ means that the summation is taken over the nearest-neighbor sites. 
An actual calculation is 
performed as follows. Hereafter, we mainly focus our discussion on the half-filled case, unless 
otherwise stated.  Equations 
(\ref{eq:R1}) and (\ref{eq:R2}) are diagonalized by means of a unitary transformation ${\cal U}$ 
to give the mean-field Hamiltonian 
\begin{equation}
H_{\rm mf}=\sum_{m=1}^{N_{\rm L}}\varepsilon_{m}\gamma^{\dagger}_{m\uparrow}\gamma_{m\uparrow}
+\sum_{m=1}^{N_{\rm L}}(-\varepsilon_{m})\gamma^{\dagger}_{m\downarrow}\gamma_{m\downarrow},
\label{eq:R3}
\end{equation}
where $N_{\rm L}$ is the number of lattice sites, $0\le\varepsilon_{1}\le\varepsilon_{1}\dots
\le\varepsilon_{N_{\rm L}}$, and the fermion operators $\gamma$ describing the quasiparticles 
are related to the electron operators $c$ by 
\begin{eqnarray}
& &[c^{\dagger}_{1\uparrow}, \cdots,\,c^{\dagger}_{N_{\rm L}\uparrow},\,
c_{1\downarrow}, \cdots,\,c_{N_{\rm L}\downarrow}]
\nonumber 
\\
& &\qquad\qquad
=[\gamma^{\dagger}_{1\uparrow},\cdots,\, \gamma^{\dagger}_{N_{\rm L}\uparrow},\, 
\gamma_{1\downarrow},\cdots,\, \gamma_{N_{\rm L}\downarrow}]{\cal U}^{\dagger}. 
\label{eq:R4}
\end{eqnarray}
Substituting Eq.\ (\ref{eq:R4}) into Eq.\ (\ref{eq:CS2}), we obtain the self-consistent 
equation for the gap $\Delta_{ij}$ as 
\begin{eqnarray}
& &
\Delta_{ij}=\frac{V}{2}\sum_{m=1}^{N_{\rm L}}\left[
({\cal U})^{*}_{j+N_{\rm L},m}({\cal U})_{i,m}
-({\cal U})^{*}_{i+N_{\rm L},m}({\cal U})_{j,m}\right]
[1-f(\varepsilon_{m})]
\nonumber
\\
& &
\qquad
+\frac{V}{2}\sum_{m=1}^{N_{\rm L}}\left[
({\cal U})^{*}_{j+N_{\rm L},m+N_{\rm L}}({\cal U})_{i,m+N_{\rm L}}
-({\cal U})^{*}_{i+N_{\rm L},m+N_{\rm L}}({\cal U})_{j,m+N_{\rm L}}\right]
f(\varepsilon_{m}), 
\label{eq:R5}
\end{eqnarray}
where ${\cal U}$ depends on $\Delta_{ij}$ and $\varepsilon_{m}$ ($m=1,\cdots, N_{\rm L}$), 
and $f(x)$ is the Fermi distribution function $f(x)\equiv(e^{x}+1)$.    

\begin{figure}[h]
\begin{center}
\rotatebox{0}{\includegraphics[width=0.35\linewidth]{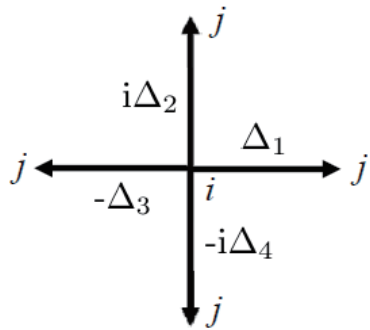}}
\caption{Phase of superconducting gap $\Delta_{ij}$ with $i$-th site chosen as a center. 
}
\label{Gap_Phase}
\end{center}
\end{figure}

We have solved Eqs.\ (\ref{eq:CS2}), (\ref{eq:CS3}), and (\ref{eq:R3})-(\ref{eq:R5}) 
self-consistently using the numerical diagonalization method and obtained the gap 
$\Delta_{ij}$ and the energy levels $\varepsilon_{m}$ ($m=1,\cdots, N_{\rm L}$).  
Numerical calculations have been performed for lattice sizes of up to 
$N_{\rm L}=30\times 30$ squares with both open and periodic boundary conditions.  
Throughout the present paper, the phase of the superconducting gap $\Delta_{ij}$ is chosen 
as shown in Fig.\ \ref{Gap_Phase}, while $\Delta_{i}$ ($i=1$-$4$) are determined 
self-consistently.  

First of all, the IMM [Eq.\ (\ref{eq:CS11})] and IAM [Eq.\ (\ref{eq:CS12})] 
are shown to be zero for the periodic boundary condition for which 
$\Delta_{1}=\Delta_{2}=\Delta_{3}=\Delta_{4}$, resulting in the recovery of the chiral gap 
given by Eq.\ (\ref{eq:CS4}) in the ${\bf k}$-space representation.   
This is because the orbital currents 
due to the relative motion of the Cooper pairs cancel with each other.  
On the other hand, if we use the open boundary condition, we obtain finite values of the IMM and 
IAM because the cancellation of the relative motion of the Cooper pairs is incomplete near 
the boundary of the system within the coherence length $\xi_{0}$ of the Cooper pairs, as in the 
case of the classical theory of magnetic materials where the current ${\bf j}_{\rm M}$ 
given by Eq.\ (\ref{eq:6}) exists only in a thin surface layer of the system if the 
magnetization ${\bf M}$ is uniform in the bulk of the system.

\begin{figure}[h]
\begin{center}
\rotatebox{0}{\includegraphics[width=0.5\linewidth]{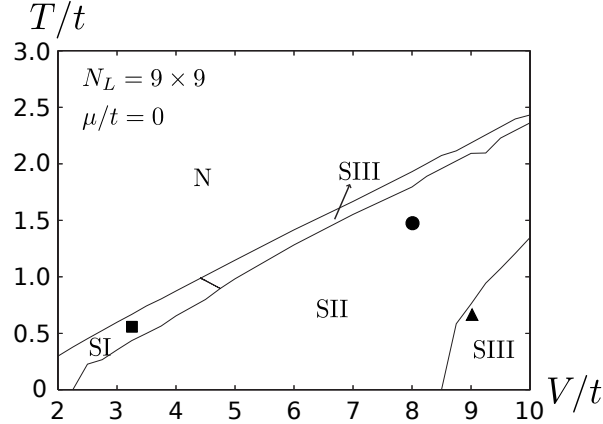}}
\caption{Phase diagram in $T/t$-$V/t$ plane. The system size is taken as $N_{\rm L}=9\times 9$. 
Three superconducting phases, SI, SII, and SIII, exist together with the normal phase N. 
}
\label{Phase_Diagram}
\end{center}
\end{figure}

Figure \ref{Phase_Diagram} shows the phase diagram in the $T/t$-$V/t$ plane for the system 
size $N_{\rm L}=9\times 9$ in the half-filled case.  Three different superconducting 
states exist, which we call SI, SII, and SIII, as shown in the figure.  
The SI phase appears around the phase boundary between the SII phase and the normal phase in the 
intermediate-coupling region $V/t<4.5$.  On the other hand, the SIII phase appears not only 
around the phase boundary between the SII phase and the normal phase in the strong-coupling 
region $V/t>4.5$, but also as a low-temperature phase in the strong-coupling region $V/t>8.5$.   

\begin{figure}[h]
\begin{center}
\rotatebox{0}{\includegraphics[width=0.5\linewidth]{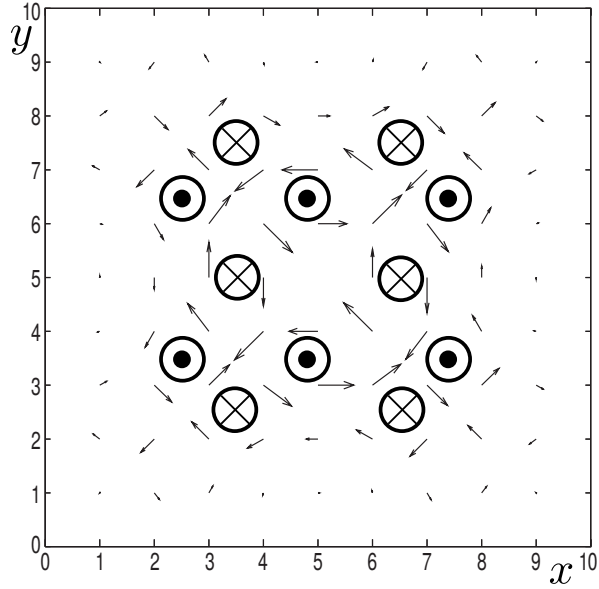}}
\caption{Pattern of current in the SI phase for $V/t=3.25$ and $T/t=0.564$ as 
indicated by a closed square in Fig.\ \ref{Phase_Diagram}.  
The system size is taken as $N_{\rm L}=9\times 9$. 
The current at each site is 
shown by an arrow whose length represents the relative size of the lattice momentum 
defined by Eq.\ (\ref{eq:CS9}).  Symbols $\otimes$ and $\odot$ indicate 
localized vortices with opposite circulations.  
}
\label{Current_SI}
\end{center}
\end{figure}

Figure \ref{Current_SI} shows the pattern of current in the SI phase for the 
parameter set, $V/t=3.25$ and $T/t=0.564$, shown by a closed square in Fig.\ \ref{Phase_Diagram}.  
The system size is taken as $N_{\rm L}=9\times 9$. 
The phases of the superconducting gap $\Delta_{ij}$ shown in Fig.\ \ref{Gap_Phase} are given by 
$\Delta_{1}={\tilde \Delta}_{1}$, $\Delta_{2}=-{\rm i}{\tilde \Delta}_{2}$, 
$\Delta_{3}=-{\tilde \Delta}_{3}$, and $\Delta_{4}={\rm i}{\tilde \Delta}_{4}$, where 
${\tilde \Delta}_{1}\sim {\tilde \Delta}_{4}$ are real and have the same sign.  Note that 
the sign of ${\tilde \Delta}_{1}\sim{\tilde \Delta}_{4}$ is determined so as to satisfy 
Eq.\ (\ref{eq:CS2a}). 
In this phase, the current is induced around the center of the bulk and its pattern can be seen 
as a deformed vortex pair lattice as shown by symbols $\otimes$ and $\odot$.  
The physical reason why this pattern is realized is unclear 
for the moment.{~\cite{VSauls}}

\begin{figure}[h]
\begin{center}
\rotatebox{0}{\includegraphics[width=0.5\linewidth]{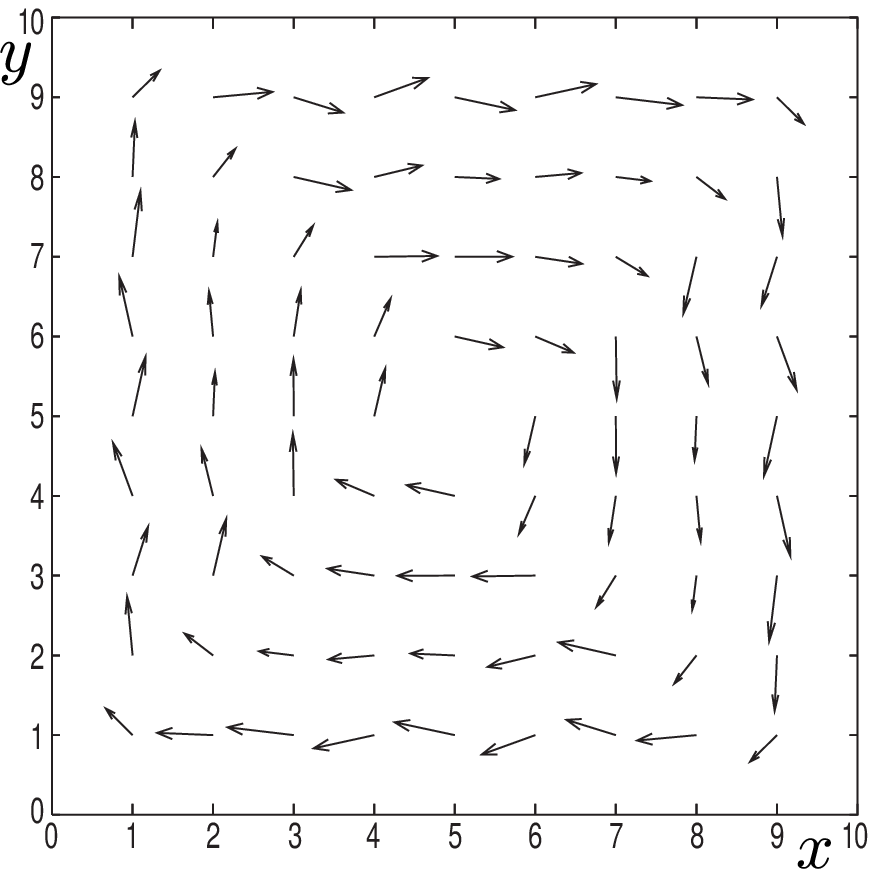}}
\caption{Pattern of current in the phase SII for $V/t=8.00$ and $T/t=1.482$ as 
indicated by a closed circle in Fig.\ \ref{Phase_Diagram}.  
The system size is taken as $N_{\rm L}=9\times 9$.  Current at each site is 
shown by an arrow whose length represents the relative size of lattice momentum 
defined by Eq.\ (\ref{eq:CS9}).
}
\label{Current_SII}
\end{center}
\end{figure}

Figure \ref{Current_SII} shows the pattern of current in the SII phase for 
the parameter set, $V/t=8.00$ and $T/t=1.482$, shown by a closed circle in 
Fig.\ \ref{Phase_Diagram}.  The system size is taken as $N_{\rm L}=9\times 9$. 
This phase is the bulk phase in the intermediate-coupling region as shown in 
Fig.\ \ref{Phase_Diagram}.  
The phases of the superconducting gap $\Delta_{ij}$ shown in Fig.\ \ref{Gap_Phase} are given by 
$\Delta_{1}\sim\Delta_{4}$, which are real and have the same sign. 
The pattern of the current distribution is that expected physically.  Namely, the current is induced 
near the boundary of the system owing to incomplete cancellation of the relative angular momentum of 
the Cooper pairs.  The reason why the current exists at the center of the system is that the 
system size $N_{\rm L}=9\times 9$ is comparable to the extent of the Cooper pair 
$\xi^{*}(T)\equiv \pi\xi(T)$, $\xi(T)$ being the coherence length at finite $T$.

\begin{figure}[h]
\begin{center}
\rotatebox{0}{\includegraphics[width=0.5\linewidth]{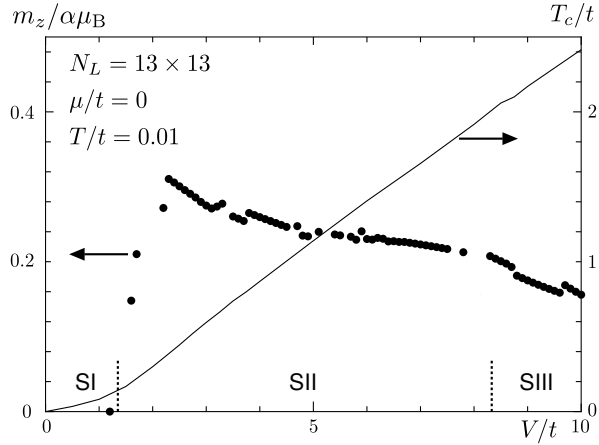}}
\caption{IMM/$\mu_{0}$ per site $m_{z}$ as a function of the attractive interaction 
$V/t$ for the system size $N_{\rm L}=13\times 13$ at $T/t=0.01$ in the case of half-filling, i.e., 
$\mu/t=0$. SI, SII, and SIII indicate the phases shown in Fig.\ \ref{Phase_Diagram}.
The dimensionless parameter $\alpha$ is defined by $\alpha\equiv tma^{2}/\hbar^{2}$. 
}
\label{m_z_1}
\end{center}
\end{figure}

If we follow the BCS theory for an s-wave weak-coupling superconductor, the 
coherence length in the ground state (at $T=0$) is given by $\xi_{0}=\hbar v_{\rm F}/\pi\Delta$.  
Then, the ratio of $\xi_{0}$ to the lattice constant $a$ is estimated as 
\begin{equation}
\frac{\xi_{0}}{a}=\frac{\gamma\hbar v_{\rm F}}{a\pi^{2}T_{\rm c}}
\simeq\frac{2\gamma}{\pi^{3}}\frac{E_{\rm F}}{T_{\rm c}}
\simeq0.11\frac{E_{\rm F}}{T_{\rm c}},
\label{CoherenceLength}
\end{equation}
where we have used the BCS relation $\Delta=\pi T_{\rm c}/\gamma$, 
where $\log\,\gamma$ is the Euler constant $C=0.57721\cdots$, 
and assumed the free dispersion 
for the quasiparticles and that the Fermi momentum is given by $p_{\rm F}\simeq \hbar \pi/a$.  
As shown later in Fig.\ \ref{m_z_1}, $T_{\rm c}/t\simeq 1.8$ for the attractive interaction 
$V/t=8.00$.  The Fermi energy $E_{\rm F}$ is estimated as $E_{\rm F}\simeq 4t$ in the case of 
half-filling.  Therefore, with the use of Eq.\ (\ref{CoherenceLength}), $\xi_{0}/a$ is 
estimated as $\xi_{0}/a\simeq 0.24$, giving the estimated extent of the Cooper 
pair $\xi^{*}=\pi\xi_{0}$ as $\xi^{*}/a\simeq0.75$.  
On the other hand, the temperature $T/t=1.482$  for 
Fig.\ \ref{Current_SII} is about 90\% of the transition temperature $T_{\rm c}$, i.e., 
$(T_{\rm c}-T)/T_{\rm c}\simeq 0.1$, as seen in Fig.\ \ref{Phase_Diagram}.  Then, with the use of 
the correlation length $\xi(T)\simeq 0.74\xi_{0}\sqrt{T_{\rm c}/(T_{\rm c}-T)}$ at a finite 
temperature $T$,\cite{deGennes3} the extent of the Cooper pair at $T/t=1.482$ is estimated as 
$\xi^{*}(T)/a\simeq 0.75\times 0.74\times \sqrt{10}\simeq 1.76$, which is not negligible
compared with the system size $N_{\rm  L}=9\times 9$.  

\begin{figure}[h]
\begin{center}
\rotatebox{0}{\includegraphics[width=0.7\linewidth]{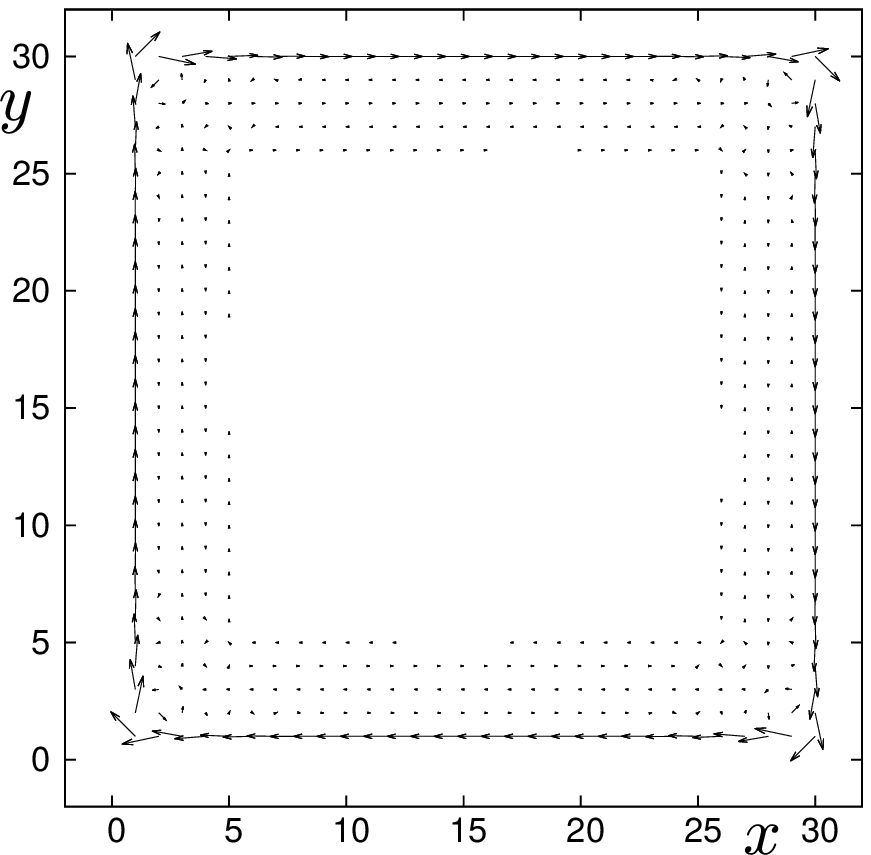}}
\caption{Pattern of current in the SII phase for $V/t=8.00$ and $T/t=1.482$.  
The system size is taken as $N_{\rm L}=30\times 30$.  The current at each site is 
shown by an arrow whose length represents the relative size of the lattice momentum 
defined by Eq.\ (\ref{eq:CS9}).  
}
\label{Current_SII_Large}
\end{center}
\end{figure}

Figure \ref{Current_SII_Large} shows the pattern of current in the SII phase for 
the same parameter set as above, $V/t=8.00$ and $T/t=1.482$, but for a much larger system size, 
$N_{\rm L}=30\times 30$.  We can see that the current is essentially confined near the system 
boundary with width $\xi^{*}\simeq 1.76a$. 
The current along and near the boundary of the system can be regarded as a lattice version of 
the surface current of the A-phase of $^3$He, which is a topological superfluid characterized by the 
topological class D defined in Ref.\ ~\citen{Schnyder}.

\begin{figure}[h]
\begin{center}
\rotatebox{0}{\includegraphics[width=0.5\linewidth]{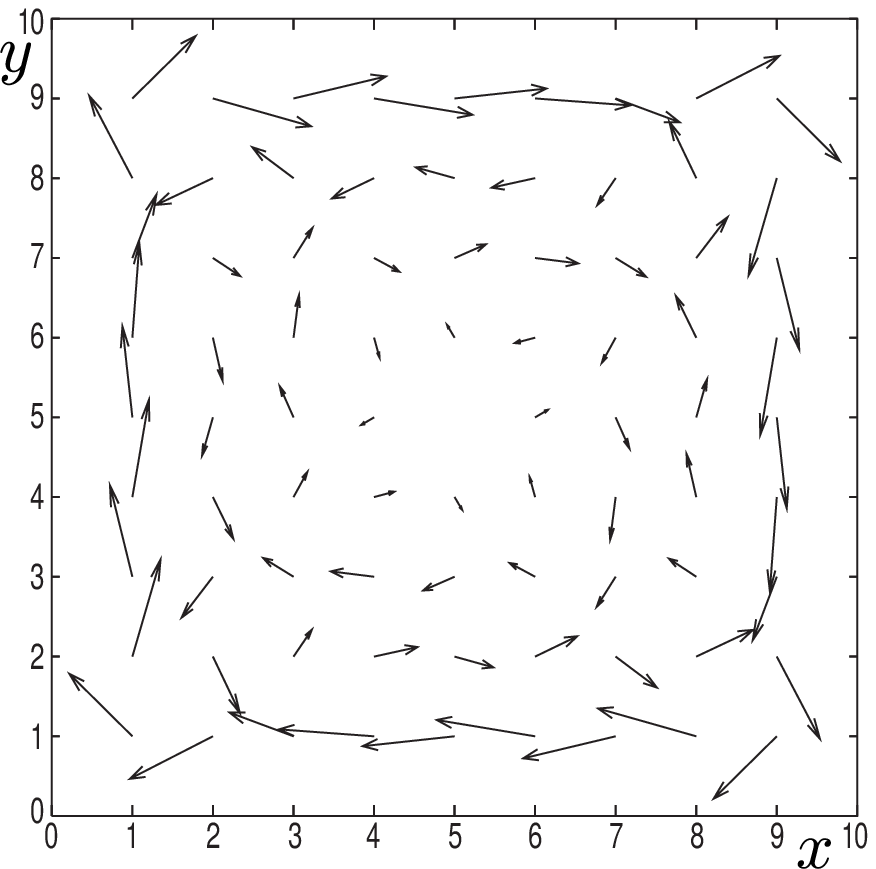}}
\caption{Pattern of current in the SIII phase for $V/t=9.00$ and $T/t=0.644$ as 
indicated by a closed triangle in Fig.\ \ref{Phase_Diagram}.  
The system size is taken as $N_{\rm L}=9\times9$.  The current at each site is 
shown by an arrow whose length represents the relative size of the lattice momentum 
defined by Eq.\ (\ref{eq:CS9}).
}
\label{Current_SIII}
\end{center}
\end{figure}

Figure \ref{Current_SIII} shows the pattern of current in the SIII phase for 
the parameter set, $V/t=9.00$ and $T/t=0.644$, shown by a closed triangle in 
Fig.\ \ref{Phase_Diagram}. 
The system size is taken as $N_{\rm L}=9\times9 $. 
The phases of the superconducting gap $\Delta_{ij}$ shown in Fig.\ \ref{Gap_Phase} are given by 
$\Delta_{1}\sim\Delta_{4}$ which are real and have the same sign. In this phase, the induced current 
forms concentric layers of flows, with adjacent layers having opposite signs. 
The physical reason why this pattern is realized is unclear for the moment.

In Fig.\ \ref{m_z_1}, $m_z\equiv M_{z}/\mu_{0}N_{\rm L}$, which is 
the IMM per site divided by $\mu_{0}$, and $T_{\rm c}/t$, the superconducting transition 
temperature, are shown as functions of$ V/t$, which is the strength of the attractive 
interaction, at $T/t=0.01$ 
for the system size $N_{\rm L}=13\times 13$ at half-filling ($\mu/t=0$).  
Note that, in order to avoid the effect of the boundary, $T_{c}/t$ is calculated with the 
periodic boundary condition.  
It is noteworthy that $m_z$ does not decrease even though $T_{\rm c}$ decreases and that 
a negative correlation exists between $m_z$ and $T_{c}/t$ except in the SI phase 
where the coherence length becomes comparable to 
the system size and the superconducting state is greatly suppressed over the whole system. 
This suggests that the IAM, connected to the IMM by Eqs.\ (\ref{eq:CS11}) and 
(\ref{eq:CS12}), is on the order of $\hbar N/2$ without the extra factor 
$(T_{\rm c}/T_{\rm F})^{\gamma}$ ($\gamma=1$ or 2).  

Indeed, according to 
Eqs.\ (\ref{eq:CS11}) and (\ref{eq:CS12}), the IAM, $L_{\rm in}$, is expressed in terms of 
$\hbar N$ and $m_{z}/\alpha \mu_{\rm B}$ as follows: 
\begin{equation}
L_{\rm in}=\frac{2m_{\rm b}}{(-e)}\frac{M_{z}}{\mu_{0}}
=-\frac{\alpha m_{\rm b}}{m}\hbar N_{\rm L}\frac{m_{z}}{\alpha \mu_{\rm B}},
\label{eq:R6}
\end{equation}   
where $\alpha\equiv tma^{2}/\hbar^{2}$ and 
we have used the definition of the Bohr magneton $\mu_{\rm B}=e\hbar/2m$.  
Since it is easily derived that $\alpha m_{\rm b}/m=1/2$ if we use the definition of $\alpha$ 
given above  and Eq.\ (\ref{eq:CS10}), the IAM 
at half-filling (i.e., $N_{\rm L}=N$) is given by 
\begin{equation}
L_{\rm in}=-\frac{\hbar}{2}N\frac{m_{z}}{\alpha \mu_{\rm B}}.
\label{eq:R7}
\end{equation}   
Considering the size of $m_{z}/\alpha \mu_{\rm B}$ in the SII phase shown in Fig.\ \ref{m_z_1}, 
we can see that the IAM in the SII phase is given by $L_{\rm in}\sim\hbar N/2$.   
This implies that $\gamma=0$ for the exponent of the extra factor $(T_{\rm c}/T_{\rm F})^{\gamma}$, 
verifying the validity of the original theory by Ishikawa for the IAM 
in the $^{3}$He-A phase.~\cite{Ishikawa}

\begin{figure}[h]
\begin{center}
\rotatebox{0}{\includegraphics[width=0.5\linewidth]{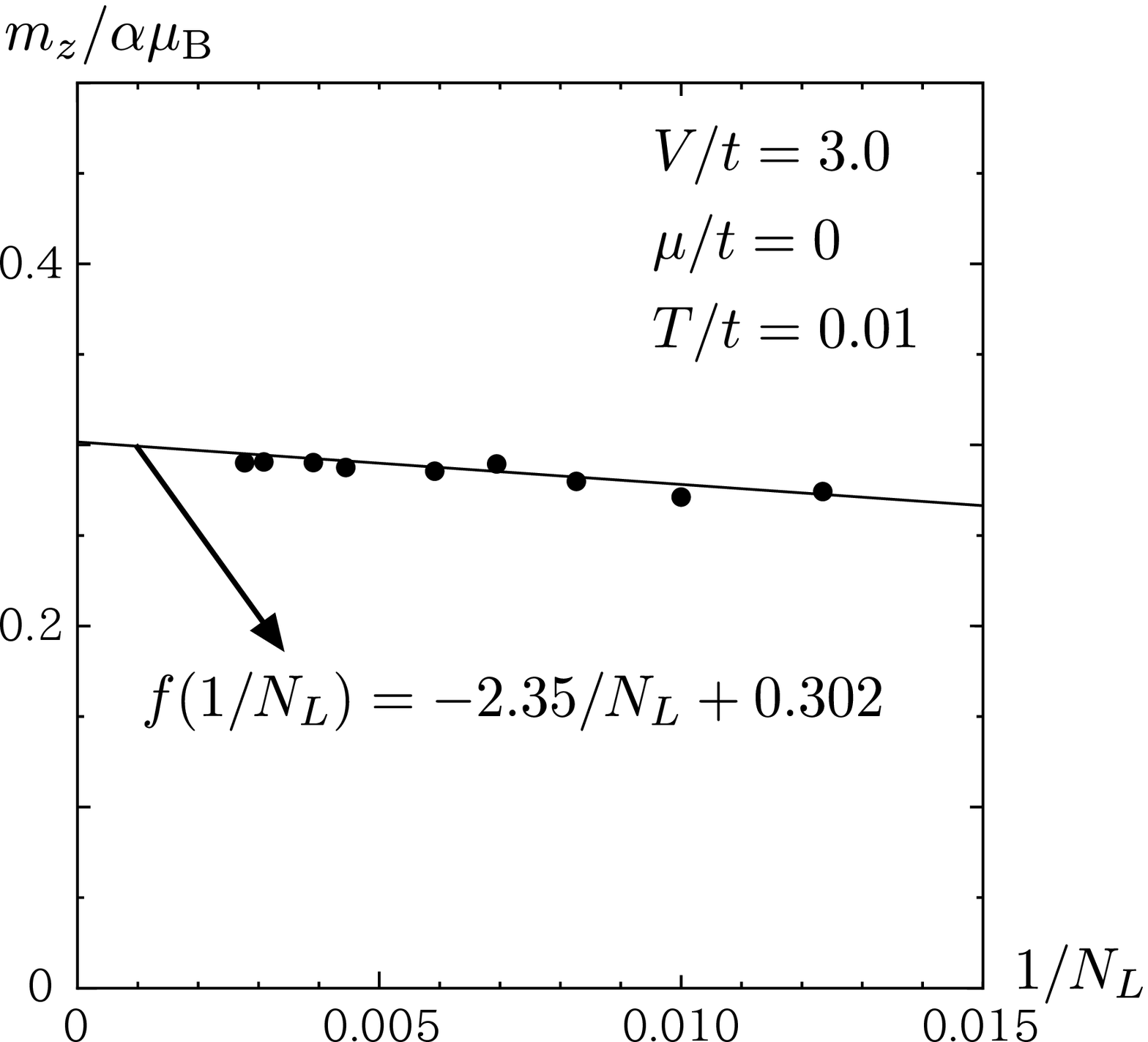}}
\caption{System-size scaling for IMM/$\mu_{0}$ per site $m_{z}$ 
for $V/t=3.0$ and $T/t=0.01$ at half-filling, i.e.,  $\mu/t=0$. 
The dimensionless parameter $\alpha$ is defined by $\alpha\equiv tma^{2}/\hbar^{2}$.
}
\label{m_z_2}
\end{center}
\end{figure}

Figure \ref{m_z_2} shows the system-size dependence of $m_{z}/\alpha \mu_{\rm B}$ for the 
case of $V/t=3.0$ and $T/t=0.01$ at half-filling, i.e., $\mu/t=0$, up to $N_{\rm L}=19\times 19$.  
We can see that the system-size scaling works reasonably well, giving 
$m_{z}/\alpha \mu_{\rm B}\simeq 0.302$ in the limit $N_{\rm L}\to \infty$.  

\begin{figure}[h]
\begin{center}
\rotatebox{0}{\includegraphics[width=0.5\linewidth]{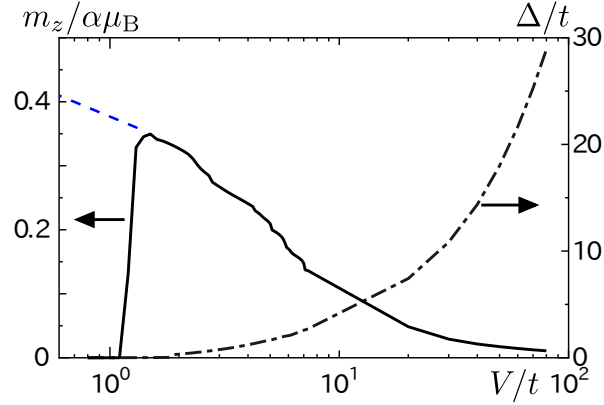}}
\caption{IMM/$\mu_{0}$ per site $m_{z}$ and magnitude of the superconducting gap $\Delta$ 
as functions of the attractive interaction 
$V/t$ for the system size $N_{\rm L}=30\times 30$ at $T/t=0$ at half-filling, i.e.,  
$\mu/t=0$. 
The dimensionless parameter $\alpha$ is defined by $\alpha\equiv tma^{2}/\hbar^{2}$.  
The dashed curve is that expected in the limit $N_{\rm L}\to \infty$. 
}
\label{No:11}
\end{center}
\end{figure}

Figure \ref{No:11} shows how IMM/$\mu_{0}$ per site, $m_{z}/\alpha \mu_{\rm B}$, changes 
depending on the strength of the attractive interaction $V/t$, together with 
the behavior of $\Delta/t$, with $\Delta=\Delta_{1}=\Delta_{2}=\Delta_{3}=\Delta_{4}$ defined 
in Fig.\ \ref{Gap_Phase}.  The system size is 
taken as $N_{\rm L}=30\times 30$, the maximum size adopted in the present paper.  
The reason why $m_{z}/\alpha\mu_{\rm B}$ approaches zero at $V/t\simeq 1$ can be understood 
as follows.  At $V/t\simeq 1$, the transition temperature is $T_{\rm c}/t\simeq 0.08$ as shown 
in Fig.\ \ref{m_z_1}.  Therefore, according to $\xi_{0}/a$ given by Eq.\ (\ref{CoherenceLength}) 
and $\xi^{*}=\pi\xi_{0}$, the extent of the Cooper pair $\xi^{*}/a$ for 
$T_{\rm c}/t\simeq0.08$ is estimated as  
\begin{equation}
\frac{\xi^{*}}{a}=\frac{\pi\xi_{0}}{a}\simeq\pi\times0.11\times\frac{4t}{0.08t}\simeq17.
\label{CoherenceLength3}
\end{equation}
This implies that the superconducting order for the system size $N_{\rm L}=30\times 30$ is 
almost destroyed by the effect of the boundary of the system.  

\begin{figure}[h]
\begin{center}
\rotatebox{0}{\includegraphics[width=0.6\linewidth]{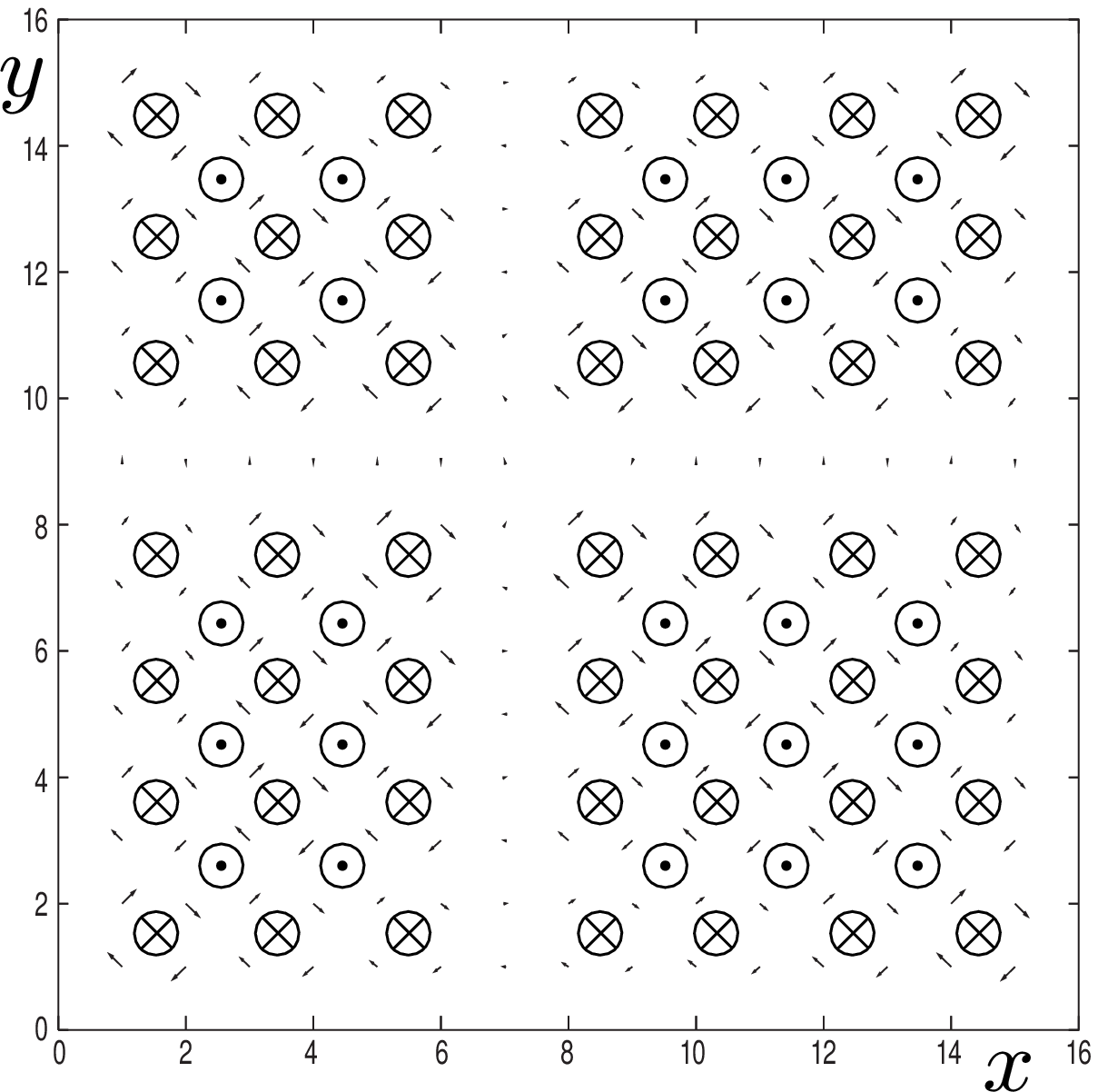}}
\caption{
Pattern of current in the SIII phase for $V/t=50$ and $T/t=0$.  
The system size is taken as $N_{\rm L}=15\times15$. 
The current at each site is shown by an arrow whose length represents the relative size of 
the lattice momentum defined by Eq.\ (\ref{eq:CS9}).  Symbols $\otimes$ and $\odot$ indicate 
localized vortices with opposite circulations.  
}
\label{No:12}
\end{center}
\end{figure}

We can see that 
$m_{z}/\alpha\mu_{\rm B}$ decreases as $V/t$ increases.  The $V/t$ dependence of 
$m_{z}/\alpha\mu_{\rm B}$ in the region $V/t>10$ can be fitted by const./($V/t$), although 
we do not show this explicitly.  In this region, the pattern of the current corresponds to the 
lattice of vortices and antivortices with domain walls as shown in Fig.\ \ref{No:12} for 
the system size $N_{\rm L}=15\times 15$ with $V/t=50$ and $T/t=0$.  Namely, the tendency that 
the magnetic moments of the vortex and 
antivortex cancel with each other becomes prominent in this region.  However, this extremely 
strong coupling region does not seem to be realized in actual systems. 

The dashed curve in Fig.\ \ref{No:11} is a smooth extrapolation to $V/t\to 0$ where 
$m_{z}/\alpha\mu_{\rm B}$ is 
expected to be $m_{z}/\alpha\mu_{\rm B}=4/\pi^{2}$, as argued below.  
Indeed, according to Eq.\ (\ref{eq:10}) and Eq.\ (\ref{A:4}), which is valid at half-filling, 
and the definition of $\alpha$, $\alpha\equiv tma^{2}/\hbar^{2}$, $m_{z}/\alpha\mu_{\rm B}$ is 
transformed as follows: 
\begin{eqnarray}
& &
\frac{m_{z}}{\alpha\mu_{\rm B}}=\frac{1}{2}\frac{m}{m_{\rm band}^{\rm occ}}
\frac{\mu_{\rm B}}{\alpha\mu_{\rm B}}=
\frac{1}{2\alpha}\frac{4}{\pi^{2}}\frac{m}{m_{\rm b}}
\nonumber
\\
& &
\qquad
=\frac{4}{\pi^{2}}=0.40528\cdots\simeq 0.41,
\label{eq:R8}
\end{eqnarray}
where, in deriving the last equality, we have used the relation $\alpha m_{\rm b}/m=1/2$, which is 
derived from Eq.\ (\ref{eq:CS10}).  

\begin{figure}[h]
\begin{center}
\rotatebox{0}{\includegraphics[width=0.5\linewidth]{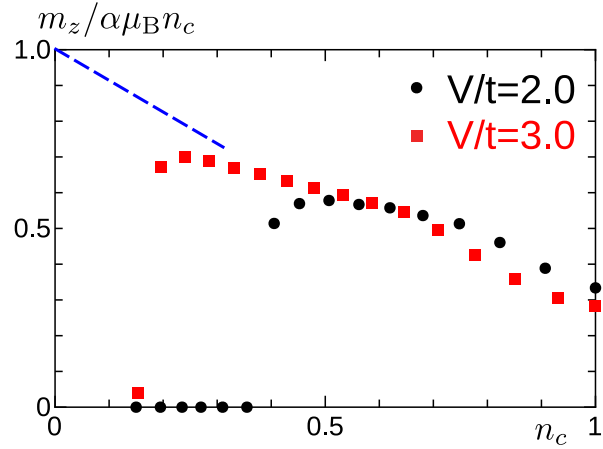}}
\caption{(Color online) IMM/$\mu_{0}$ per particle $m_{z}/\alpha\mu_{\rm B}n_{c}$ as a function of 
the filling $n_{c}$ for the attractive interactions, $V/t=2.0$ and 3.0.  
The system size is taken as $N_{\rm L}=30\times 30$.  
The dimensionless parameter $\alpha$ is defined by $\alpha\equiv tma^{2}/\hbar^{2}$.  
The dashed curve is that expected in the limit $n_{c}\to 0$. 
}
\label{No:13}
\end{center}
\end{figure}

Figure \ref{No:13} shows the filling ($n_{c}$) dependence of IMM/$\mu_{0}$, 
$m_{z}/\alpha\mu_{\rm B}n_{c}$, for the attractive interactions $V/t=2.0$ and 3.0 
at $T/t=0$.  The reason why $m_{z}/\alpha\mu_{\rm B}n_{c}$ approaches zero at 
$n_{c}\simeq 0.15$ in the case of $V/t=3.0$ can be understood by the effect of competition 
between the system 
size and the extent of the Cooper pairs as in the case in Fig.\ \ref{No:11}.  
The extent of the Cooper pairs $\xi^{*}$ is estimated by Eq.\ (\ref{CoherenceLength}) 
but with $a$ replaced by $a/\sqrt{n_{c}}$ because 
the average distance between electrons increases in inverse proportion to the square root 
of the filling $n_{c}$. Then, the following relation holds:
\begin{equation}
\frac{\xi^{*}}{a}\simeq \pi\times 0.11\frac{E_{\rm F}}{T_{\rm c}\sqrt{n_{c}}}.  
\label{eq:R9}
\end{equation}    
The transition temperature $T_{\rm c}$ is calculated as $T_{\rm c}\simeq 0.071t$.  
Therefore, $\xi^{*}$ for $V/t=3.0$ and the filling $n_{c}=0.15$, 
which gives $E_{\rm F}\simeq 0.887t$, 
is estimated as 
\begin{equation}
\frac{\xi^{*}}{a}\simeq 11, 
\label{eq:R10}
\end{equation}
which is approximately half of the size of the system of 30$a$.  
This explains why $m_{z}$ becomes zero at approximately $n_{c}=0.15$.  

The dashed curve in Fig.\ \ref{No:13} is a smooth extrapolation to $n_{c}\to 0$, where 
$m_{z}/\alpha\mu_{\rm B}n_{c}$ is expected to be equal to 1.  
Indeed, in the dilute limit ($n_{c}\to 0$), the effect of the lattice fades away so that 
the result in free space should be recovered.  The IAM $L_{\rm in}$ in the free space is 
expected to be given by 
$L_{\rm in}=\hbar N/2$.  On the other hand, extending Eq.\ (\ref{eq:R7}), 
$L_{\rm in}$ is given by 
\begin{equation}
L_{\rm in}=-\frac{\hbar}{2}N\frac{m_{z}}{\alpha\mu_{\rm B}n_{c}},
\label{eq:R12}
\end{equation}
where $n_{c}=N/N_{\rm L}$.  Therefore, in the limit $n_{\rm c}\to 0$, 
${m_{z}}/{\alpha\mu_{\rm B}n_{c}}$ is expected to approach 1.

\newpage
\section{Effect of Meissner Current on IAM}
\begin{figure}[h]
\begin{center}
\rotatebox{0}{\includegraphics[width=0.9\linewidth]{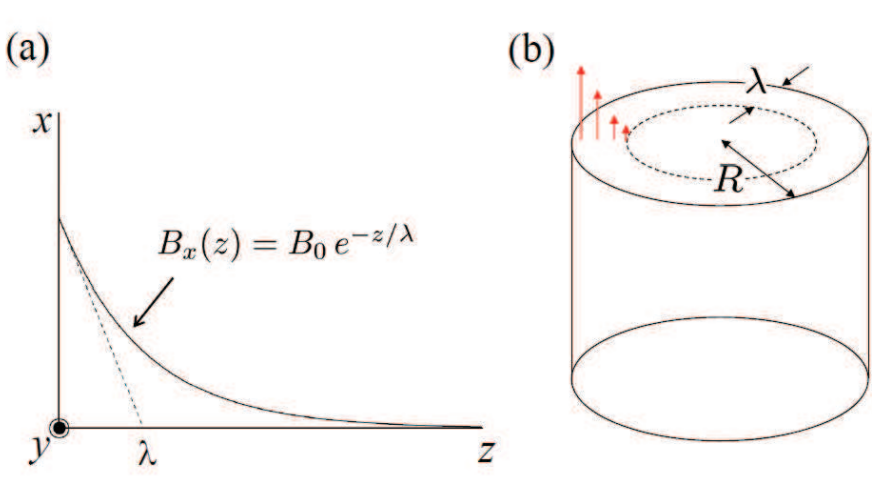}}
\end{center}
\caption{(Color online)
(a) Magnetic field near the boundary between a superconductor and a vacuum. 
(b) Distribution of the magnetic field (indicated by arrows) in a cylindrical superconductor 
sample.
 }
\label{Fig:ME1}
\end{figure}

In this section, we discuss the effect of the Meissner current on the IMM and IAM induced in 
a chiral superconductor as discussed in the previous section.  
Here, we estimate the orbital angular momentum due to the Meissner current flowing in a 
thin surface layer within penetration depth $\lambda$.  First of all, let us consider the 
situation shown in Fig.\ \ref{Fig:ME1}(a), where the flat boundary surface of the superconductor is 
the $xy$-plane and the magnetic field $B_{x}(z)$ is parallel to the $x$-direction and decreases 
in the superconductor ($z>0$).  Then, by the Amp\`ere law, the current density $j_{y}(z)$ 
is given by 
\begin{equation}
j_{y}(z)={1\over \mu_{0}}{dB_{x}(z)\over dz}.
\label{eq:ME1}
\end{equation}
Therefore, the Meissner current $J_{\rm M}$ (per unit length), flowing in the $y$-direction 
in a surface layer in the $x$-direction, is given by 
\begin{equation}
J_{\rm M}=\int_{0}^{\infty}dz\,{1\over \mu_{0}}{dB_{x}(z)\over dz}=-{B_{0}\over \mu_{0}},
\label{eq:ME2}
\end{equation}
where $B_{0}\equiv B_{x}(0)$.  The corresponding mass current $P_{\rm M}$ is estimated as 
\begin{equation}
P_{\rm M}\simeq {m_{\rm band}^{\rm FS}\over e}J_{\rm M},
\label{eq:ME3}
\end{equation}
where $m_{\rm band}^{\rm FS}$ is the band mass averaged near the Fermi level.  
The reasons why the mass 
$m_{\rm band}^{\rm FS}$ appears are that the actual current is caused by a deformation of the 
Fermi surface, and that the Fermi liquid correction by ``$F_{1}^{\rm s}$" due to the back flow effect 
does not cancel the dynamical mass enhancement in the case where the Galilean invariance is broken 
in lattice systems such as Sr$_2$RuO$_4$.~\cite{Leggett1,Leggett2,Varma}.  
Then, in the situation shown in Fig.\ \ref{Fig:ME1}(b), 
the orbital angular momentum $L_{\rm M}$ due to this $P_{\rm M}$ is given by 
\begin{equation}
L_{\rm M}=2\pi R\times R\times P_{\rm M}.
\label{eq:ME4}
\end{equation}
With the use of Eqs.\ (\ref{eq:ME2}) and (\ref{eq:ME3}), $L_{\rm M}$ is reduced to 
\begin{equation}
L_{\rm M}=2\pi R^{2}\times{m_{\rm band}^{\rm FS}\over e}\times\left(-{B_{0}\over \mu_{0}}\right).
\label{eq:ME5}
\end{equation}
If we equate $B_{\rm in}=M_{\rm in}$ in Eq.\ (\ref{eq:12}) to $B_{0}$ at the surface, 
$L_{\rm M}$ is finally given by 
\begin{equation}
L_{\rm M}=-{n(\pi R^{2})\hbar\over 2}\,
{m_{\rm band}^{\rm FS}\over m_{\rm band}^{\rm occ}}
=-{N\hbar\over 2}\,{m_{\rm band}^{\rm FS}\over m_{\rm band}^{\rm occ}}.
\label{eq:ME6}
\end{equation}
The first equality of Eq.\ (\ref{eq:ME6}) is also valid for columnar systems with a general 
cross-section shape if the factor $\pi R^{2}$ is replaced by the cross-section area $S$ 
because the angular momentum is related to the areal velocity.   
In this sense, the expression for $L_{\rm M}$, Eq.\ (\ref{eq:ME6}), is valid for columnar 
systems with an arbitrary shape of the cross section.   
This result implies that $L_{\rm M}$ cannot cancel the IAM $L_{\rm in}=N\hbar/2$ in general 
as far as $m_{\rm band}^{\rm FS}\not= m_{\rm band}^{\rm occ}$, which is usually the case 
including the case of free space, as for liquid $^{3}$He, 
if the many-body effect near the Fermi surface is taken 
into account.  Namely, the total angular momentum $L$ in the ground state, given by 
\begin{equation}
L=L_{\rm in}+L_{\rm M}=\frac{N\hbar}{2}
\left(1-\frac{m_{\rm band}^{\rm FS}}{m_{\rm band}^{\rm occ}}\right),
\label{eq:ME7}
\end{equation}
is not vanishing but is on the order of ${\cal O}(N\hbar)$ and has the opposite sign to 
the ${\hat{\ell}}$ 
vector because $m_{\rm band}^{\rm FS}>m_{\rm band}^{\rm occ}$ in correlated systems in general.  
We note, however, that the total angular momentum $L$, given by Eq.\ (\ref{eq:ME7}), vanishes in the 
hypothetical system with a free electron dispersion because 
$m_{\rm band}^{\rm FS}/m_{\rm band}^{\rm occ}=1$ in such a case.  

\section{Multi-band Effect}
\begin{figure}[h]
\begin{center}
\rotatebox{0}{\includegraphics[width=0.5\linewidth]{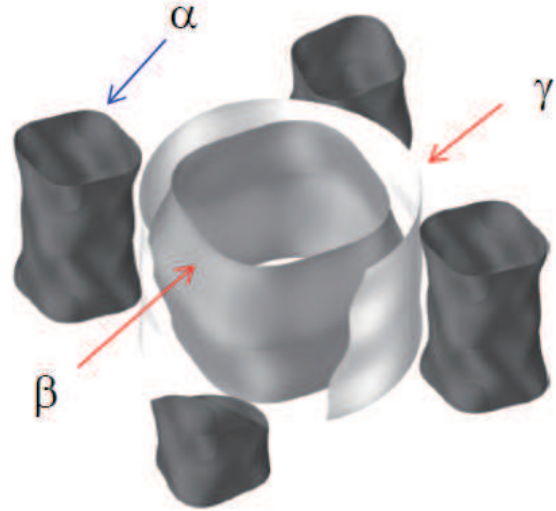}}
\end{center}
\caption{(Color online)
Fermi surfaces of Sr$_2$RuO$_4$. The $\alpha$-band is hole-like, 
and the $\beta$- and $\gamma$-bands are electron-like.   
 }
\label{Fig:MB1}
\end{figure}

The discussions so far has been for a single-band model. However, multiple bands exist in general. 
In particular, Sr$_2$RuO$_4$, which is a promising candidate for exhibiting the IAM and IMM, 
has three bands, one hole-like band ($\alpha$) and two electron-like bands ($\beta$ and $\gamma$) 
(see Fig.\ \ref{Fig:MB1}).    
Therefore, the IMMs of the electron-like and hole-like bands partially cancel with each other 
because they have opposite signs, reflecting the opposite signs of the gyromagnetic ratio.  
The characters of the bands are summarized in 
Table\ \ref{Table:1}.~\cite{Mackenzie2}  
According to Eq.\ (\ref{eq:10}), the IMM is inversely proportional to the band mass.  
Therefore, the relative size of the IMM is also inversely proportional to the band mass.  
The relative value of the IMM is given in the last row in Table\ \ref{Table:1} 
for the case that the IAMs of each 
band have the same magnitude.  Then, total IMM density, $M_{\rm in}^{\rm tot}$, is 
approximately given by 
\begin{equation}
M_{\rm in}^{\rm tot}\simeq -{n_{\rm s}\over 2}\mu_{0}\mu_{\rm B}
\left(-\frac{1}{1.1}+\frac{1}{2.2}+\frac{1}{2.9}\right),
\label{eq:MB1}
\end{equation}
if Eq.\ (\ref{eq:10}) is valid for the three bands shown in Table\ \ref{Table:1}.  
The factor in parenthesis in Eq.\ (\ref{eq:MB1}) gives a small value of $-0.110$; 
thus,  the magnitude of 
$M_{\rm in}^{\rm tot}$ is one order smaller than that expected in the single-band model:
\begin{equation}
M_{\rm in}^{\rm tot}\simeq {n_{\rm s}\over 2}\mu_{0}\mu_{\rm B}
\times 0.110.
\label{eq:MB2}
\end{equation}
Therefore, $M_{\rm in}^{\rm tot}$ induces the magnetic field 
$B_{\rm in}\simeq 2.3\times 10^{-3}\,{\rm T}=23\,{\rm G}$, which is smaller than 
the lower critical field $B_{{\rm c}1}^{\rm obs}\simeq 5.0\times 10^{-3}\,{\rm T}$ and 
is easily screened out by the Meissner effect, in contrast to the case of the single-band model 
as discussed in Sect\ 1.2.

\begin{table}[tb]
\caption{Band-dependent $m_{\rm band}/m$ according to Ref.\ \citen{Mackenzie2} and 
the IAM and IMM ratios.}
\label{Table:1}
\begin{tabular}{cccc}
\hline
Band & $\alpha$ & $\beta$ & $\gamma$ \\
Character & Hole & Electron & Electron \\
\cline{1-4}
$m_{\rm band}/m$ & 1.1 & 2.2 & 2.9 \\
IAM ratio & 1 & 1 & 1\\
IMM ratio & 1 & $-$0.5 & $-$0.38 \\ 
\hline
\end{tabular}
\end{table}

On the other hand, this cancellation does not occur in the total IAM, $L_{\rm in}^{\rm tot}$; 
$L_{\rm in}^{\rm tot}$ is given by the sum of the IAM of three bands, which are all expected 
to be on the order of ${\cal O}(N_{\rm s}\hbar/2)$. 
Therefore, the effect of the Meissner current on the total IAM is much less important than 
in the case of the single-band IAM given by Eq.\ (\ref{eq:ME7}). 
Thus, the total IAM remains to be technically 
unscreened by the Meissner current.  This multiband effect is crucial to distinguish 
the IAM from the IMM.

\section{How to Observe IMM and IAM in Sr$_2$RuO$_4$}
In this section we discuss how to detect the IMM and IAM by carrying out experiments that are 
possible to perform on Sr$_2$RuO$_4$. 
 
As discussed in the previous sections, it will be difficult to detect the magnetic field 
induced by the IMM owing to the Meissner screening in the bulk sample. 
However, it may be possible to detect it for a small sample with size on the order of 
the penetration depth $\lambda\sim 1300\,$\AA.  
On the other hand, the current pattern near the boundary of a small sample will be 
rather complicated in the sense that the directions of the Meissner current and the current 
inducing the IMM are opposite and the ranges of both currents are different, leading to 
considerable cancellation between the two currents up to the penetration depth $\lambda$. 
Therefore, in order to detect the IMM, the experiments should be carefully designed. 

Another possibility exists of observing the IMM by performing $\mu$SR experiment, 
which can detect the IMM induced around the stopping site of $\mu^{+}$ in principle.  
This is because $\mu^{+}$ attracts electrons from adjacent 
Ru sites and acts as a nonmagnetic impurity that destroys the chiral superconducting order at 
sites surrounding $\mu^{+}$, resulting in a local circulating current around the $\mu^{+}$ 
site because the cancellation of the chiral current of Cooper pairs becomes incomplete there.  
If the effective impurity potential is sufficiently strong to completely suppress the chiral order 
at surrounding sites,~\cite{MiyakeNarikiyo} the magnetic field induced by this circulating current 
will be on the order of $B_{\rm in}$ [Eq.\ (\ref{eq:13})], which can be easily 
shown by standard calculations in classical electrodynamics based on the expressions for the 
surface current given by Eqs. \ (\ref{eq:5}) and (\ref{eq:7}).  A crucial point here is that 
$B_{\rm in}$ at the $\mu^{+}$ site is free from the Meissner screening effect.  
On the other hand, if the effective impurity potential is not so strong, 
the induced magnetic field at the $\mu^{+}$ site is decreased considerably 
depending on the strength of the potential.  Indeed, the magnetic field observed by 
$\mu$SR, $B_{\mu{\rm SR}}^{\rm obs}\simeq 0.5\,{\rm G}$, is rather small compared with 
$B_{\rm in}\simeq 2.3\times 10^{-3}\,{\rm T}=23\,{\rm G}$ given by Eq.\ (\ref{eq:MB2}). 
This fact can be understood by assuming that the effective impurity potential from $\mu^{+}$ 
is only moderate.  However, this should be verified by an explicit model calculation, 
which is left for a future study. 

On the other hand, the IAM of each band gives additive contributions to the total IAM on the 
order of ${\cal O}(N_{\rm s}\hbar)$ which can be probed by the so-called 
Richardson-Einstein-de Haas effect, which has been used to detect the macroscopic spin 
angular momentum in the ferromagnetic state.~\cite{Einstein,Richardson}  
Since the size of the IAM is on the same order as the spin angular momentum of 
ferromagnetic compounds, it is expected that the IAM can be observed by this effect in practice, 
although a sufficiently low temperature will be required. 

\section{Conclusion}
On the basis of the tight-binding model with the nearest-neighbor attraction on a square lattice, 
the IMM and IAM have been calculated by solving the Bogoliubov-de Gennes equation. 
It turned out that, in the ground state, the IMM $m_{z}$ per site divided by $\mu_{0}$ is 
on the order of $\mu_{\rm B}$ and the IAM is on the order of $L_{\rm in}\sim \hbar N$.  
In particular, in the dilute limit, the result strongly indicates that the IAM approaches 
$L_{\rm in}=\hbar N/2$, which is the value first predicted by Ishikawa.~\cite{Ishikawa} 
The IAM and IMM are induced by the surface current flowing in a thin layer with a width 
on the order of the coherence length $\xi_{0}$. 

It has been shown that thus created IAM is partially 
screened by the Meissner current if the particles have an electric charge. The extent of 
the cancellation depends on the ratio of the effective mass $m_{\rm band}^{\rm FS}$ 
of quasiparticles near the Fermi level to that of the harmonic average $m_{\rm band}^{\rm occ}$ 
over the occupied state: 
if they were equal, the cancellation would become perfect.  This is in marked 
contrast to the IMM, which is almost completely screened by the Meissner effect if 
the spontaneous magnetic field created by the IMM is smaller than the lower critical field 
$H_{{\rm c}1}$, as in the case of Sr$_2$RuO$_4$.  

On the other hand, it turned out that the multiband effect is important, as in the case of 
Sr$_2$RuO$_4$.  Namely, a certain amount of cancellation of the IMM occurs among particle and 
hole bands because they have charges with different signs, while such a cancellation does not 
occur for the IAM.  

An interpretation of the spontaneous magnetic field observed in Sr$_2$RuO$_4$ by $\mu$SR, and 
a possible means of probing the IAM have also been proposed.  

\section*{Acknowledgments}
We are grateful to M. Akatsu and T. Goto for directing our attention to the 
Richardson-Einstein-de Haas effect as a possible means of observing the intrinsic 
angular momentum of Sr$_2$RuO$_4$. Communication with S. Kashiwaya on the detectability of 
the intrinsic magnetic moment in Sr$_2$RuO$_4$ is also acknowledged. 
This work is supported by a Grant-in-Aid for Scientific 
Research on Innovative Areas ``Topological Quantum Phenomena" 
(No.22103003) from the Ministry of Education, Culture, Sports, Science and
Technology of Japan, and by a Grant-in-Aid for Scientific 
Research (No.25400369) from Japan Society for the Promotion of Science.

\newpage
\appendix
\section{Harmonic Average of Band Mass of Electrons on Square Lattice}
\begin{figure}[h]
\begin{center}
\rotatebox{0}{\includegraphics[width=0.6\linewidth]{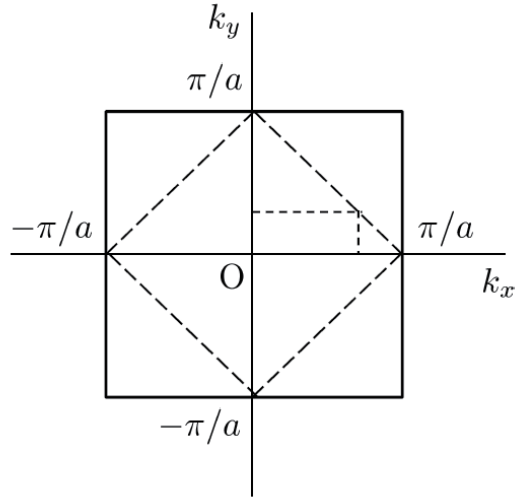}}
\end{center}
\caption{
Brillouin zone of electrons on the square lattice. Dashed lines indicate the Fermi surface 
at half-filling and dotted lines indicate the order of two-dimensional integration.  
}
\label{Fig:A1}
\end{figure}

In this appendix, we derive the harmonic average of the band mass of electrons 
with dispersion, Eq.\ (\ref{eq:CS13}).  The inverse mass tensor ${\hat m}^{-1}$ 
of a band electron is given by 
\begin{equation}
{\hat m}^{-1}={1\over \hbar^{2}}\left(
\begin{array}{cc}
\displaystyle {\partial^{2}\epsilon_{k}\over\partial k_{x}^{2}} & 
\displaystyle {\partial^{2}\epsilon_{k}\over\partial k_{x}\partial k_{y}}\\
\displaystyle {\partial^{2}\epsilon_{k}\over\partial k_{y}\partial k_{x}} & 
\displaystyle {\partial^{2}\epsilon_{k}\over\partial k_{y}^{2}}
\end{array}
\right).
\label{A:1}
\end{equation}
Substituting the dispersion of an electron, Eq.\ (\ref{eq:CS13}), into this expression, 
it is easily seen that the matrix in Eq.\ (\ref{A:1}) is diagonal.  Indeed, its explicit form is 
\begin{equation}
{\hat m}^{-1}={2ta^{2}\over \hbar^{2}}\left(
\begin{array}{cc}
\cos\,k_{x}a & 0 \\
0 & \cos\,k_{y}a
\end{array}
\right).
\label{A:2}
\end{equation}
The averaging of ${\hat m}^{-1}_{xx}$ over the occupied states at half-filling 
is performed as follows: 
\begin{equation}
\langle{\hat m}^{-1}_{xx}\rangle={2ta^{2}\over \hbar^{2}}{1\over {1\over 2}(\pi/a)^{2}}
\int_{0}^{\pi/a}dk_{y}\int_{0}^{\pi/a-k_{y}}dk_{x}\,\cos\,k_{x}a
={2ta^{2}\over \hbar^{2}}\times{4\over \pi^{2}},
\label{A:3}
\end{equation}
where the area of integration with respect to $k_{x}$ and $k_{y}$ is restricted to the 
part surrounded by the dashed line (the Fermi surface) of the first quadrant in Fig.\ \ref{Fig:A1}.  
The averaging of ${\hat m}^{-1}_{yy}$ is performed in a similar way, giving the same value.  
Then, using the definition of the band mass (near the $\Gamma$-point) 
$m_{\rm b}$ [Eq.\ (\ref{eq:CS10})], the harmonic average of the band mass $m_{\rm band}^{\rm occ}$ 
over the occupied states is given by 
\begin{equation}
m_{\rm band}^{\rm occ}={1\over \langle{\hat m}^{-1}_{xx}\rangle}={\pi^{2}\over 4}m_{\rm b}.
\label{A:4}
\end{equation}

\section{Effective Mass Averaged over the Fermi Surface}
In this appendix, we derive the effective mass averaged over the Fermi surface 
of electrons with dispersion, Eq.\ (\ref{eq:CS13}).  
The inverse mass tensor ${\hat m}^{-1}$ of this band electron is given by Eq.\ (\ref{A:2}).  
Therefore, the average of ${\hat m}^{-1}_{xx}$ over the Fermi surface is calculated as follows: 
\begin{eqnarray}
& &\langle{\hat m}^{-1}_{xx}\rangle={2ta^{2}\over \hbar^{2}}
\frac{\displaystyle \int d{\bf k}\cos\,k_{x}a\,
\delta(\cos\,k_{x}a+\cos\,k_{y}a+\mu/2t)}
{\displaystyle \int d{\bf k}\,\delta(\cos\,k_{x}a+\cos\,k_{y}a+\mu/2t)}
\nonumber
\\
& &\qquad\quad
={1\over 2}\,{2ta^{2}\over \hbar^{2}}
\frac{\displaystyle \int d{\bf k}\,(\cos\,k_{x}a+\cos\,k_{y}a)
\delta(\cos\,k_{x}a+\cos\,k_{y}a+\mu/2t)}
{\displaystyle \int d{\bf k}\,\delta(\cos\,k_{x}a+\cos\,k_{y}a+\mu/2t)}
\nonumber
\\
& &\qquad\quad
=-{1\over 2}{\mu a^{2}\over \hbar^{2}}.
\label{B:1}
\end{eqnarray}
The expression for $\langle{\hat m}^{-1}_{yy}\rangle$ is also given by Eq.\ (\ref{B:1}).  
Therefore, the effective mass averaged over the Fermi surface $m_{\rm band}^{\rm FS}$ is 
given by 
\begin{equation}
m_{\rm band}^{\rm FS}=-{2\hbar^{2}\over\mu a^{2}}.
\label{B:2}
\end{equation}
It is remarked that $m_{\rm band}^{\rm FS}$ diverges toward half-filling, i.e., 
$\mu\to 0$, which is consistent with the existence of the van Hove singularity in the 
density of states at $\mu=0$.  On the other hand, in the dilute limit, 
i.e., $\mu\to -4t$, $m_{\rm band}^{\rm FS}$ 
approaches $\hbar^{2}/2t a^{2}$, which is the same as the band mass $m_{\rm b}$ near the 
$\Gamma$-point defined by Eq.\ (\ref{eq:CS10}).  This guarantees the validity of the definition 
in Eq.\ (\ref{B:2}).  In the almost-filled case, $\mu\simeq 4t$, $m_{\rm band}^{\rm FS}$ 
approaches $-\hbar^{2}/2t a^{2}=-m_{\rm b}$, the hole band mass near the top of the band.

\end{document}